\documentclass[aps,prl,onecolumn,10pt,superscriptaddress,nofootinbib,nobibnotes,longbibliography]{revtex4-2}

\usepackage{amsmath,amssymb,amsthm,mathtools,mathrsfs}
\usepackage{graphicx}
\usepackage{xcolor}
\usepackage[
  colorlinks=true,
  linkcolor=magenta,
  citecolor=blue,
  urlcolor=magenta,
  filecolor=magenta,
  linktoc=all,
  pdfborder={0 0 0}
]{hyperref}

\newtheorem{mainresult}{Theorem}
\newtheorem{theorem}{Theorem}[section]
\newtheorem{lemma}[theorem]{Lemma}
\newtheorem{proposition}[theorem]{Proposition}
\newtheorem{corollary}[theorem]{Corollary}
\theoremstyle{plain}
\newtheorem{predecessortheorem}{Theorem}
\newtheorem{predecessorconjecture}{Conjecture}
\theoremstyle{definition}
\newtheorem{definition}[theorem]{Definition}
\theoremstyle{remark}

\newcommand{\C}{\mathbb C}
\newcommand{\Lop}{\mathcal L}
\newcommand{\Tr}{\operatorname{Tr}}
\newcommand{\CPTP}{\operatorname{CPTP}}
\newcommand{\id}{\mathbb I}
\newcommand{\ketket}[1]{|#1\rangle\!\rangle}
\newcommand{\ketketbra}[2]{|#1\rangle\!\rangle\langle\!\langle #2|}

\newcommand{\ket}[1]{\lvert #1\rangle}
\newcommand{\bra}[1]{\langle #1\rvert}
\newcommand{\brabra}[1]{\langle\!\langle #1|}
\newcommand{\SW}{\mathsf S}
\newcommand{\Span}{\operatorname{span}}
\newcommand{\Ad}{\operatorname{Ad}}

\newcommand{\diag}{\operatorname{diag}}

\begin{document}
\raggedbottom
\clubpenalty=10000
\widowpenalty=10000
\displaywidowpenalty=10000
	
\title{Simulating the quantum switch with quantum circuits is impossible}

\author{Caisheng Cheng}
\email{ccs112202@mail.ustc.edu.cn}
\affiliation{Hefei National Laboratory for Physical Sciences at Microscale and Department of Modern
	Physics, University of Science and Technology of China, Hefei, Anhui, China}
\affiliation{Shanghai Branch, CAS Centre for Excellence and Synergetic Innovation Centre in Quantum Information and Quantum Physics, University of Science and Technology of China, Shanghai 201315, China}
\affiliation{Shanghai Research Center for Quantum Sciences, Shanghai 201315, China}

\begin{abstract}
The quantum switch is a higher-order transformation that combines input
quantum channels in an indefinite causal order. Whether quantum circuits
can simulate its action, and how many calls to the input channels such
a simulation requires, are natural questions about the power of quantum
control of causal order. In this work, we prove that no quantum circuit
with classically controlled causal order can exactly and deterministically
simulate the quantum switch for arbitrary input channels using any finite
number of calls. This impossibility holds already for qubit channels and
extends to the complete $r$-switch, which coherently controls all
permutations of $r$ input channels. These results settle the question of exact deterministic circuit
simulation of the quantum switch. They demonstrate that processes with
indefinite causal order cannot, in general, be efficiently simulated in
this setting even by quantum circuits with classically controlled causal
order.

\end{abstract}

\maketitle

\setcounter{tocdepth}{2}
\begingroup
\hypersetup{linkcolor=blue}
\tableofcontents
\endgroup
\newpage

\section*{Introduction}
Quantum information processing can change the resources required for a task
and the performance attainable with those resources. Quantum advantage over
classical approaches is a central motivation for research on factoring, search,
and other computational tasks~\cite{shor1997polynomial,grover1996fast,boyer1998searching,montanaro2016overview,farhi1998decision,childs2003exponential,bernstein1997quantum,simon1997power,harrow2009linear,brassard2002amplitude,hallgren2007pell,magniez2011walk,ambainis2010formula,durr2006graphs,szegedy2004markov,clader2013preconditioned,childs2010algebraic,ambainis2012variable,kallaugher2024streaming,berry2014differential}, as well as on
quantum simulation~\cite{feynman1982simulating,lloyd1996universal,jordan2012field,lanyon2011digital,trotzky2012relaxation,georgescu2014simulation,berry2015nearly,hastings2015chemistry,buluta2009simulators,blatt2012ions,aspuruguzik2012photonic,somaroo1999simulations,poulin2015trotter,trivedi2024analogue},
learning~\cite{huang2021information,aharonov2022measurement,chen2022quantummemory,huang2022experiments,chen2024pauli,oh2024bosoniclearning,liu2025photonic,wiebe2012datafitting,gilboa2024distributed,schmidhuber2025planted}, and precision
measurement~\cite{giovannetti2006metrology,giovannetti2004enhanced,zhou2018heisenberg,ligo2011shotnoise,ganapathy2023broadband,degen2017sensing,caves1981noise,wineland1992squeezing,kitagawa1993squeezed,hosten2016measurement,aasi2013enhanced,higgins2007phase,leibfried2004spectroscopy,esteve2008squeezing,leroux2010cavity,cox2016deterministic,arrad2014sensing,kessler2014metrology,dur2014metrology,unden2016metrology,davis2016heisenberg}. Information processing also
depends on how operations are combined, which makes quantum control of their
order a possible source of advantage.

A quantum control can preserve coherence between different sequences of
operations, rather than fixing a sequence or selecting one classically~\cite{procopio2015orders,rubino2017verification,goswami2018switch,wei2019switching,guo2020transmission,rozema2024experimental,antesberger2024tomography}.
Protocols with indefinite causal order offer advantages in channel
discrimination~\cite{chiribella2012perfect,bavaresco2021strict,bavaresco2022unitary},
metrology~\cite{zhao2020quantum,mothe2024reassessing,liu2023optimal,yin2023indefinite}, computational
complexity~\cite{araujo2014computational,renner2022computational,taddei2021multiple}, quantum query
complexity~\cite{abbott2024quantum}, and transformations of unknown unitary
and isometry operations~\cite{quintino2022deterministic,yoshida2025universal,quintino2019inverting,yoshida2023decoders}.
Several of these advantages are realized with the quantum
switch~\cite{PhysRevA.88.022318,Bavaresco2025}. It sends the same target through
two unknown channels in opposite orders under the coherent control of a
qubit. The complete $r$-switch extends this action to all permutations of
$r\geq2$ input channels, using each once and retaining coherence between
the orders~\cite{PRXQuantum.2.030335}. These switches take channels as inputs
and produce a joint channel on the control and target.

The discrimination and metrology advantages are established at specified
numbers of channel calls.
Allowing additional calls might let a quantum circuit reproduce the switch's
entire joint output channel. If it could do so with only a modest increase
in calls, it could replace the switch wherever that operation is used.
The simulating circuit may choose the order of channel calls using classical
information. Such circuits are called quantum circuits with classical
control of causal order (QC--CC). They retain quantum processing and memory,
while classical outcomes determine which channel is called
next~\cite{PRXQuantum.2.030335}. The possibility of replacing coherent
control with repeated access motivates a broader simulation question.
Can additional calls allow these circuits to reproduce higher-order
transformations with indefinite causal order for arbitrary unknown input
channels?

In the ideal limit of unlimited access to the input channels, the answer
is yes. Process tomography would determine their descriptions and allow
the prescribed output operation to be
implemented~\cite{poyatos1997complete,chuang1997prescription}.
Exact characterization requires unboundedly many calls, because finite
measurement statistics leave uncertainty about an arbitrary unknown
channel. For the switch, another affirmative answer is available if
unsuccessful runs can be discarded. Postselection produces the required coherent
action on successful outcomes rather than on every
run~\cite{PhysRevA.88.022318,milz2018entanglement}.
The question is thus whether both costs can be avoided by one scheme that
uses finitely many calls and reproduces the switch on every run for
arbitrary unknown input channels.

For unitary inputs, one extra call to one of the two channels suffices for
an exact deterministic simulation of the two-channel
switch~\cite{PhysRevA.88.022318,Bavaresco2025}. For arbitrary channels,
however, Bavaresco \emph{et al.} showed that if one input is called only
once, the other requires exponentially many calls in the number of qubits
per input channel, even for adaptive QC--CC
simulation~\cite{Bavaresco2025}. This exponential separation is an important
advance. The theorem's restriction to a single call to one input leaves
open whether this separation persists when both input channels can be
queried repeatedly~\cite{Bavaresco2025}. Could the separation be stronger
still, with no finite number of calls sufficient for an exact deterministic
universal QC--CC simulation?

In this work, we establish an infinite query separation between quantum
and classical control of causal order in the exact universal simulation
of the switch. For every fixed target dimension $d\geq2$, exact
deterministic universal QC--CC simulation of the complete two-channel
switch is impossible with any fixed finite number of calls to each input,
whereas coherent switching uses each input once. This separation persists
for the complete coherent $r$-switch, which controls all permutations of
$r\geq2$ input channels. Its exact deterministic universal QC--CC simulation
is impossible with any fixed finite total number of calls, whereas
coherent switching uses $r$ calls.

\newpage

\section*{Our results}
\emph{Quantum switch simulation.}---The two-channel quantum switch takes
arbitrary unknown channels $\mathcal A,\mathcal B\in\CPTP_d$ to a channel
$\mathsf S_d^{(2)}(\mathcal A,\mathcal B)$ on a qubit control and a
$d$-dimensional target~\cite{PhysRevA.88.022318}. Here $\CPTP_d$ denotes the
completely positive, trace-preserving maps on a $d$-dimensional system,
with $d\geq2$. The control state $|0\rangle$ sends the target through
$\mathcal A$ then $\mathcal B$, while $|1\rangle$ reverses this order.
Choose Kraus representations $\{A_i\}$ and $\{B_j\}$ of $\mathcal A$
and $\mathcal B$, respectively. With the control as the first tensor
factor, the switched channel has Kraus operators $S_{ij}$ and acts on an
arbitrary joint control-target input state $\rho$ as
\begin{equation*}
\begin{aligned}
S_{ij}
= |0\rangle\!\langle0|\otimes B_jA_i
 + |1\rangle\!\langle1|\otimes A_iB_j,\qquad
\mathsf S_d^{(2)}(\mathcal A,\mathcal B)(\rho)
= \sum_{i,j}S_{ij}\rho S_{ij}^{\dagger}.
\end{aligned}
\end{equation*}
The off-diagonal control blocks contain cross terms between the two
ordered products, describing interference between the orders.

To reproduce this channel, consider a QC--CC circuit making exactly $M$
calls to $\mathcal A$ and $N$ to $\mathcal B$, with $M,N$ predetermined
nonnegative integers. Repeated calls to each input access the same
unknown memoryless channel. Before and between calls, the circuit
processes the system together with arbitrary finite private quantum
memory. These operations are quantum instruments whose classical
outcomes select the next unused call. Both the instruments and the call
choices may depend on earlier outcomes~\cite{PRXQuantum.2.030335}.
The complete circuit, including its final quantum operation, induces a
deterministic channel $\mathcal C_{M,N}[\mathcal A,\mathcal B]$ on the joint
control and target when all internal outcomes are summed. For fixed
$d,M,N$, one circuit, including its internal operations
and their conditional rules, must be chosen independently of the unknown
channels. Exact universal simulation requires this same circuit to
satisfy, for every $\mathcal A,\mathcal B\in\CPTP_d$,
\begin{equation*}
\mathcal C_{M,N}[\mathcal A,\mathcal B]
=\mathsf S_d^{(2)}(\mathcal A,\mathcal B).
\end{equation*}
This channel equality also includes inputs entangled with an arbitrary
reference system.

\begin{mainresult}[Two-channel switch]\label{thm:main_two_channel}
For every fixed integer $d\geq2$ and every finite pair
$M,N\in\mathbb Z_{\geq0}$, no deterministic QC--CC circuit using $M$ calls
to $\mathcal A$ and $N$ calls to $\mathcal B$ exactly reproduces the complete
two-channel switch $\mathsf S_d^{(2)}(\mathcal A,\mathcal B)$ for all
$\mathcal A,\mathcal B\in\CPTP_d$.
\end{mainresult}

We give a proof sketch here; the detailed statement and proof appear in
Theorem~\ref{thm:supp_all_finite_binary} and
Section~\ref{sec:supp_finite_binary}.
Repeated calls can produce higher-order dependence on the unknown
channels, whereas the switch is linear in each input channel separately.
This suggests isolating the part of a proposed simulator that uses each
input once. For a qubit simulator with $M,N>0$, retain one call to each
input and fill all other calls with the identity channel $\mathcal I$.
The known calls can be absorbed into the circuit's internal operations.
Choosing the retained pair uniformly at random then gives a deterministic
QC--CC circuit $\overline{\mathcal C}$ using each unknown input once.

The uniform average is determined by the original simulation equality.
Indeed, apply that equality to the physical mixtures
$(\mathcal I+t\mathcal A)/(1+t)$ and
$(\mathcal I+s\mathcal B)/(1+s)$, with $s,t\geq0$.
After multiplication by $(1+t)^M(1+s)^N$, the coefficient of $ts$
on the circuit side sums over all retained pairs. Comparing it with the
switch side gives
\begin{equation*}
\overline{\mathcal C}[\mathcal A,\mathcal B]
=\mathsf S_2^{(2)}\!\left(
\left(1-\frac1M\right)\mathcal I+\frac{\mathcal A}{M},\,
\left(1-\frac1N\right)\mathcal I+\frac{\mathcal B}{N}\right).
\end{equation*}
Thus the average realizes a specific mixture of the complete switch,
the two single-channel operations, and the identity, with switch weight
$1/(MN)$ (Corollary~\ref{cor:positive-average}).
Identity filling also gives these terms a useful structure: whenever
either input is replaced by the identity, the two orders coincide.
The three other paths therefore act identically in both control sectors,
so comparing those sectors can isolate the switched contribution.

To exploit this structure, consider unitary input channels with matrices
$U,V$. The required mixture then has four Kraus amplitudes, and
positivity forces every Kraus amplitude of the circuit to lie in their
span. Group the circuit histories
by which input is called first and fix a Kraus decomposition of the
internal operations in each group. Inserting matrices $X,Y$ in the two
calls gives a bilinear operator amplitude $G_{o\alpha}(X,Y)$, where
$o=A,B$ labels the order and $\alpha$ labels the history and internal
Kraus choices.
The span condition gives an expansion for each chosen pair of inputs.
To compare these expansions with causal normalization, we need
coefficients fixed by the circuit across all inputs.
Lemma~\ref{lem:physical} establishes such a decomposition:
\begin{equation*}
\begin{aligned}
G_{o\alpha}(X,Y)
=e_{o\alpha}w(X,Y)+\id_C\otimes H_{o\alpha}(X,Y),\qquad
w(X,Y)
=|0\rangle\!\langle0|\otimes YX+
|1\rangle\!\langle1|\otimes XY.
\end{aligned}
\end{equation*}
Here each $e_{o\alpha}$ is a fixed scalar and each $H_{o\alpha}$ a fixed
bilinear map, valid for all $2\times2$ matrices $X,Y$.
The lemma further resolves $H_{o\alpha}$ into the three remaining paths:
the target passes through either input separately or through neither.
Their coefficients are fixed linear or bilinear forms in the input
matrices. This structure will also determine which residual terms
survive the causal constraints.

The difference between the two diagonal control blocks of an amplitude
removes $H_{o\alpha}$ and leaves $e_{o\alpha}(UV-VU)$.
The sum of these squared norms determines the switch weight. The full output
channel also fixes its cross term with either diagonal block, which
measures the interference between the switch and the residual.
Comparing both quantities with the required mixture gives, for
noncommuting unitary $U,V$,
\begin{equation*}
\sum_{o,\alpha}|e_{o\alpha}|^2=\frac1{MN},
\qquad
\sum_{o,\alpha}\overline e_{o\alpha}H_{o\alpha}(U,V)=0.
\end{equation*}

Causal order imposes a different constraint on these same amplitudes.
Tracing out the circuit's final output uses trace preservation of its
terminal operation and leaves an identity on the last queried channel's
output. This suggests contracting each ordered process with a Hermitian
operator whose partial trace over that last output vanishes: the
contraction must then be zero. We choose these operators to compare the
control sectors as well, so that the squared residual terms cancel.
Using the fixed three-path form of $H_{o\alpha}$, one contraction for
each order isolates the squared switch coefficient and the same residual
interference coefficient (Section~\ref{sec:supp_finite_binary}\,D).
The mixed control identity above, resolved into the three paths and
continued to the identity in the relevant input, makes the sum of those interference terms
vanish. Adding the two causal identities therefore gives
\begin{equation*}
0=\sum_{o,\alpha}|e_{o\alpha}|^2=\frac1{MN}>0.
\end{equation*}
Increasing the number of calls reduces the required switch weight, but
it remains positive for every finite $M,N$; the exact target and causal
normalization require incompatible values. A physical encoding extends
the contradiction to $d>2$: each extended channel is implemented on the
entire $d$-dimensional query space using one qubit call, preserving the
query counts (Proposition~\ref{prop:supp_dimension_transfer}).
A circuit omitting either input cannot reproduce the switch's dependence
on that channel (Lemma~\ref{lem:supp_omitted_binary}).

\emph{Complete $r$-switch.}---The same obstruction applies when the control
coherently selects among all orders of $r$ input channels. For arbitrary
$\mathcal A_1,\ldots,\mathcal A_r\in\CPTP_d$, with $r,d\geq2$, let
$\mathfrak S_r$ be the permutations of the $r$ labels. In control state
$|\pi\rangle$, the target encounters
$\mathcal A_{\pi(1)},\ldots,\mathcal A_{\pi(r)}$ in that order. For Kraus
families $\{A^{(\ell)}_{a_\ell}\}$ of the input channels, the multi-index
$\boldsymbol a=(a_1,\ldots,a_r)$ contains one index per channel and is
shared across the permutation terms in each controlled Kraus operator,
\begin{equation*}
W_{\boldsymbol a}
=\sum_{\pi\in\mathfrak S_r}|\pi\rangle\!\langle\pi|\otimes
A^{(\pi(r))}_{a_{\pi(r)}}\cdots A^{(\pi(1))}_{a_{\pi(1)}}.
\end{equation*}
The complete channel acts as
$\mathsf S_d^{(r)}(\mathcal A_1,\ldots,\mathcal A_r)(\rho)
=\sum_{\boldsymbol a}W_{\boldsymbol a}\rho W_{\boldsymbol a}^{\dagger}$
on arbitrary joint control-target inputs. Coherent control realizes the
complete operation using $r$ calls, one to each
input~\cite{PRXQuantum.2.030335}.

\begin{mainresult}[Complete $r$-switch]\label{thm:main_complete_switch}
For every fixed pair of integers $r,d\geq2$ and every finite vector of
query counts $\boldsymbol K=(K_1,\ldots,K_r)\in\mathbb Z_{\geq0}^r$, no
deterministic QC--CC circuit using $K_\ell$ calls to each
$\mathcal A_\ell$ exactly reproduces the complete channel
$\mathsf S_d^{(r)}(\mathcal A_1,\ldots,\mathcal A_r)$ for all tuples of
input channels in $\CPTP_d$.
\end{mainresult}

The complete-channel construction is given in
Section~\ref{sec:supp_complete_r}\,A, and
Theorem~\ref{thm:supp_all_finite_r} and its detailed proof appear in
Section~\ref{sec:supp_complete_r}\,B.
We sketch the reduction. A complete simulator must preserve the coherence
within any selected pair of control sectors. We can therefore recover
the two-channel switch inside the full permutation control.
Encode a control qubit into two permutation states differing by an
adjacent exchange of labels $i,j$, and fill the other inputs with identity
channels. The selected branches then apply $\mathcal A_i,\mathcal A_j$ in
opposite orders. A trace-preserving decoder defined on the entire output
control space returns the control qubit and preserves its coherence with
the target. The resulting channel is the complete two-channel switch
(Proposition~\ref{prop:supp_complete_pair_reduction}).

Universality requires a proposed simulator to reproduce this encoded
operation. It must query every input, since even with all other inputs
set to identity the target depends on that channel. Applying the same
encoding and decoding to the simulator, and absorbing its known identity
calls into the internal operations, therefore produces a deterministic
QC--CC simulator retaining all $K_i,K_j$ calls to the selected channels.
Their counts remain finite, contradicting
Theorem~\ref{thm:main_two_channel}.

\section*{Acknowledgments}
This initial version presents our main results and their proofs. A fuller
manuscript is being developed with collaborators and will include
applications, further physical interpretation, and additional results
in subsequent versions.

The research presented here was carried out by the human author.
We thank ChatGPT 6 Astra for assistance in checking the proof derivations
and searching the literature, and for suggestions used to polish the
writing of the \emph{Our results} section.

We recently became aware of Goldin's work~\cite{Goldin2026}, which extends
the exponential query separation established by Bavaresco
et al.~\cite{Bavaresco2025} to circuits with repeated access to both input
channels, including approximate simulation at any fixed error below
$1/2$. For exact deterministic universal simulation, we establish a
qualitatively stronger impossibility result. Already for qubit channels,
no finite number of calls can reproduce the quantum switch. This resolves
the finite-query simulation problem at fixed dimension, and the
impossibility extends to the complete $r$-switch.

C.\ C.\ is supported by the National Natural Science Foundation of China
(Grant No.\ 124B1017).

\clearpage

	\section{Quantum switch and causally ordered simulation}
	\label{sec:supp_switch_simulation}
	
	The quantum switch coherently controls the order in which two unknown channels act. We ask whether a causally ordered circuit can reproduce the same channel when it receives repeated labelled calls to those two black boxes. Equality always means equality of the resulting input--output channels, and the vector of query counts records the number of calls carrying each label.
	
	\subsection{Channels, graphical calculus, and the quantum switch}
	\label{subsec:supp_channels_switch}
	
	Let $d\geq2$, write $\Lop(H)$ for the linear operators on a finite-dimensional Hilbert space $H$, and let $\CPTP_d\coloneqq\{\mathcal E:\Lop(\C^d)\to\Lop(\C^d)\mid\mathcal E\text{ is completely positive and trace preserving}\}$ denote the $d$-dimensional channels. The two unknown channels are $\mathcal A,\mathcal B\in\CPTP_d$. The quantum switch applies $\mathcal B\circ\mathcal A$ in control sector $\lvert0\rangle$ and $\mathcal A\circ\mathcal B$ in control sector $\lvert1\rangle$, while preserving coherence between the sectors \cite{PhysRevA.88.022318,Bavaresco2025}.
	
	The past and future spaces are
	\begin{equation}
		P=P_C\otimes P_T,
		\qquad
		F=F_C\otimes F_T,
		\qquad
		P_C\simeq F_C\simeq\C^2,
		\qquad
		P_T\simeq F_T\simeq\C^d.
		\label{eq:supp_boundary_spaces}
	\end{equation}
	The input and output of each unknown channel are the $d$-dimensional target space $\C^d$.
	We compare the target and the simulator at the Choi level. In this representation, inserting a test channel into an explicitly labelled slot is a link-product contraction.
	We use the Choi convention in which the input space precedes the output space. For a linear map $\mathcal E:\Lop(H_X)\to\Lop(H_Y)$ and a fixed orthonormal basis $\{\lvert x\rangle_X\}_x$,
	\begin{equation}
		J^{\mathcal E}_{XY}
		=\sum_{x,y}\lvert x\rangle\!\langle y\rvert_X
		\otimes\mathcal E\!\left(\lvert x\rangle\!\langle y\rvert_X\right).
	\end{equation}
	Thus $\mathcal E$ is completely positive exactly when $J^{\mathcal E}\geq0$, and it is trace preserving exactly when $\Tr_YJ^{\mathcal E}_{XY}=\id_X$ \cite{JAMIOLKOWSKI1972275,CHOI1975285,watrous2018theory}. For an operator $V:H_X\to H_Y$, vectorization in the same ordering is
	\begin{equation}
		\ketket{V}_{XY}
		\coloneqq\sum_x\lvert x\rangle_X\otimes V\lvert x\rangle_X,
		\;
		J^{V(\cdot)V^\dagger}=\ketketbra{V}{V}_{XY}.
	\end{equation}
	
	For $\mathcal F:\Lop(H_1)\to\Lop(H_2)$ and $\mathcal G:\Lop(H_2)\to\Lop(H_3)$, the link product over explicitly labelled systems is
	\begin{equation}
		J^{\mathcal F}_{12}*J^{\mathcal G}_{23}
		=\Tr_2\!\left[
		\bigl(J^{\mathcal F}_{12}\otimes\id_3\bigr)
		\bigl(\id_1\otimes(J^{\mathcal G}_{23})^{T_2}\bigr)
		\right].
		\label{eq:supp_link_product}
	\end{equation}
	The partial transpose uses the basis fixed above. With this ordering,
	$J^{\mathcal G\circ\mathcal F}=J^{\mathcal F}*J^{\mathcal G}$ \cite{Chiribella_2008}.
	For Choi vectors $\ketket{Q}_{12}\in H_1\otimes H_2$ and
	$\ketket{R}_{23}\in H_2\otimes H_3$, the compatible vector link and its
	rank-one consequence are
	\begin{align}
		\ketket{Q}_{12}*\ketket{R}_{23}
		&\coloneqq
		\sum_\mu
		\bigl[(\id_1\otimes\langle\mu\rvert_2)\ketket{Q}_{12}\bigr]
		\otimes
		\bigl[(\langle\mu\rvert_2\otimes\id_3)\ketket{R}_{23}\bigr]
		\in H_1\otimes H_3,
		\nonumber\\
		\ketketbra{Q}{Q}_{12}*\ketketbra{R}{R}_{23}
		&=
		\bigl(\ketket{Q}_{12}*\ketket{R}_{23}\bigr)
		\bigl(\ketket{Q}_{12}*\ketket{R}_{23}\bigr)^\dagger.
	\end{align}
	
	We use tensor networks and graphical calculus to display the contractions in the proofs, alongside their algebraic derivations. The basic operations are collected in the diagram below, following Refs.~\cite{Wood2015Graphical,Bridgeman2017TensorNetworks}. Connected wires contract matching indices. Rectangles represent linear maps with inputs on the right and outputs on the left, while triangles represent vectors and their duals. Cups and caps denote the unnormalized vector $\sum_x\lvert x\rangle_X\otimes\lvert x\rangle_X$ and its dual in the fixed basis. An open snake therefore straightens to the identity, whereas a closed $X$ loop gives its dimension $d_X$. Transposition and entrywise complex conjugation use the same bases, with $A^\dagger=(\overline A)^T$.

\begin{equation*}
  \vcenter{\hbox{\includegraphics[width=\linewidth]{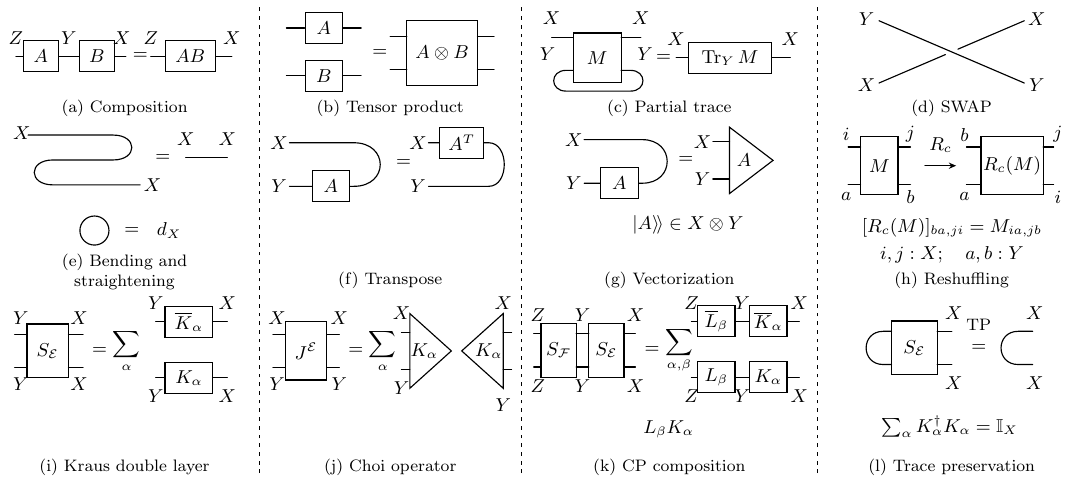}}}
\end{equation*}

For a CP map $\mathcal E(\rho)=\sum_\alpha K_\alpha\rho K_\alpha^\dagger$, the Liouville matrix $S_{\mathcal E}=\sum_\alpha\overline{K_\alpha}\otimes K_\alpha$ acts on vectorized operators as $S_{\mathcal E}\ketket{\rho}=\ketket{\mathcal E(\rho)}$. The Choi operator $J^{\mathcal E}$ instead groups its indices according to the input-first convention above. Reshuffling relates these two groupings. Each Kraus term shares its summation index across the two layers, and the trace-preserving cancellation in panel~(l) follows from the complete sum $\sum_\alpha K_\alpha^\dagger K_\alpha=\id_X$.

	Choose arbitrary Kraus representations
	$\mathcal A(\rho)=\sum_iA_i\rho A_i^\dagger$ and
	$\mathcal B(\rho)=\sum_jB_j\rho B_j^\dagger$. The switched channel is
	\begin{equation}
		\mathsf S_d^{(2)}(\mathcal A,\mathcal B)(\rho)
		=\sum_{i,j}S_{ij}\rho S_{ij}^\dagger,
		\qquad
		S_{ij}
		=\lvert0\rangle_{F_C}\!\langle0\rvert_{P_C}\otimes B_jA_i
		+\lvert1\rangle_{F_C}\!\langle1\rvert_{P_C}\otimes A_iB_j.
		\label{eq:supp_two_switch_kraus}
	\end{equation}
	Writing $\rho=\sum_{c,c'=0}^1\lvert c\rangle\!\langle c'\rvert\otimes\rho_{cc'}$, the contribution of the Kraus pair $(i,j)$ has the four control blocks
	\begin{equation}
		\begin{pmatrix}
			B_jA_i\rho_{00}A_i^\dagger B_j^\dagger
			&B_jA_i\rho_{01}B_j^\dagger A_i^\dagger\\
			A_iB_j\rho_{10}A_i^\dagger B_j^\dagger
			&A_iB_j\rho_{11}B_j^\dagger A_i^\dagger
		\end{pmatrix}.
		\label{eq:supp_two_switch_blocks}
	\end{equation}

The diagonal blocks implement the two opposite orders and the off-diagonal blocks retain their interference. Isometric changes of the two Kraus families leave the sum in Eq.~\eqref{eq:supp_two_switch_kraus} unchanged. Theorem~1 of Ref.~\cite{Dong2023quantumswitchis} shows that this extension from unitary inputs is unique among linear, completely CP-preserving two-slot supermaps.

	\subsection{Repeated labelled queries and QC--CC simulators}
	\label{subsec:supp_qccc_simulators}

	Fix exact counts $M,N\in\mathbb Z_{\geq0}$. The simulator has $M+N$ distinct open slots, with $M$ labelled by $\mathcal A$ and $N$ labelled by $\mathcal B$. All slots carrying one label are filled by the same memoryless channel. The simulator may keep arbitrary finite quantum memory between calls. It may choose the next unused slot from a classical record of earlier instrument outcomes, but distinct next-slot choices are not placed in coherent superposition \cite{PRXQuantum.2.030335}.
The physical circuit is defined for independent CP maps in its named slots, with deterministic output for independent CPTP fillings. The requirement that slots carrying the same label receive the same channel restricts the simulation task, not the physical domain of the circuit.

	For $M+N>0$, let $\Omega_{M,N}$ be the set of distinct slots. Each slot $k\in\Omega_{M,N}$ has spaces $H_k^I\simeq H_k^O\simeq\C^d$. A history $h=(k_1,\ldots,k_n)$ is an ordered list of distinct used slots. Define its set of entries by $\operatorname{used}(h)\coloneqq\{k_1,\ldots,k_n\}$, with $\operatorname{used}(\emptyset)\coloneqq\emptyset$. The initial and recursive CP elements act as
	\begin{align}
		\mathcal M_{\emptyset}^{\to k_1}
		&:\Lop(P)\longrightarrow
		\Lop\!\left(H_{k_1}^I\otimes R_{(k_1)}\right),
		&k_1&\in\Omega_{M,N},
		\nonumber\\
		\mathcal M_h^{\to k}
		&:\Lop\!\left(H_{k_n}^O\otimes R_h\right)
		\longrightarrow
		\Lop\!\left(H_k^I\otimes R_{(h,k)}\right),
		&k&\in\Omega_{M,N}\setminus\operatorname{used}(h).
	\end{align}
	Here $R_{(k_1)}$ is a finite initial memory and $h=(k_1,\ldots,k_n)$ is a nonempty history with $n<M+N$.
	For a fixed nonempty prefix $h$, let $K_h$ have orthonormal basis
	$\{\lvert k\rangle:k\in\Omega_{M,N}\setminus\operatorname{used}(h)\}$, and embed the
	outcome-dependent spaces $H_k^I\otimes R_{(h,k)}$ isometrically into their
	direct sum $\widehat H_h$ by $\iota_{h,k}$.  The instrument with its
	classical record exposed is the block-diagonal CP map
	\begin{equation}
		\widehat{\mathcal M}_h:
		\Lop(H_{k_n}^O\otimes R_h)
		\longrightarrow\Lop(K_h\otimes\widehat H_h),
		\qquad
		\widehat{\mathcal M}_h(X)
		=
		\sum_{k\in\Omega_{M,N}\setminus\operatorname{used}(h)}
		\lvert k\rangle\!\langle k\rvert_{K_h}
		\otimes
		\iota_{h,k}\mathcal M_h^{\to k}(X)\iota_{h,k}^{\dagger}.
		\label{eq:supp_classical_next_slot_register}
	\end{equation}
	For a fixed nonempty history $h$, let $M_{h,\alpha}^{\to k}:H_{k_n}^O\otimes R_h\to H_k^I\otimes R_{(h,k)}$ be Kraus operators of $\mathcal M_h^{\to k}$. On an arbitrary reference-extended input $X\in\Lop(H_{k_n}^O\otimes R_h\otimes R)$, the recorded output $Y_h\coloneqq(\widehat{\mathcal M}_h\otimes\operatorname{id}_{\Lop(R)})(X)$ has $(j,l)$ blocks

\begin{gather*}
  \vcenter{\hbox{\includegraphics{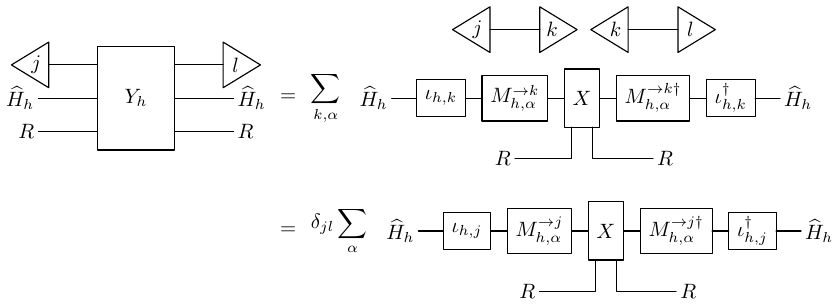}}}
  \\[4pt]
\begin{aligned}
 &(\bra j_{K_h}\otimes\id_{\widehat H_h\otimes R})Y_h(\ket l_{K_h}\otimes\id_{\widehat H_h\otimes R})\\
 &=\sum_{k,\alpha}\langle j|k\rangle\langle k|l\rangle
   (\iota_{h,k}M_{h,\alpha}^{\to k}\otimes\id_R)X
   (M_{h,\alpha}^{\to k\dagger}\iota_{h,k}^\dagger\otimes\id_R)\\
 &=\delta_{jl}\sum_\alpha
   (\iota_{h,j}M_{h,\alpha}^{\to j}\otimes\id_R)X
   (M_{h,\alpha}^{\to j\dagger}\iota_{h,j}^\dagger\otimes\id_R).
\end{aligned}
\end{gather*}
Orthogonality of the record states removes the off-diagonal blocks in $K_h$, while each diagonal block contains the corresponding conditional system--memory output, including its correlations with $R$. The initial instrument admits the same recorded direct-sum form. On every effective history input $X\geq0$, including $h=\emptyset$, normalization holds after summing over the available outcomes,

	\begin{equation}
		\sum_{k\in\Omega_{M,N}\setminus\operatorname{used}(h)}
		\Tr\!\left[\mathcal M_h^{\to k}(X)\right]
		=\Tr X.
		\label{eq:supp_qccc_instrument_normalization}
	\end{equation}
	For each complete history $h=(k_1,\ldots,k_{M+N})$, a terminal CPTP map
	$\mathcal M_h^{\to F}: \Lop\!\left(H_{k_{M+N}}^O\otimes R_h\right) \longrightarrow\Lop(F)$
	produces the global output \cite{PRXQuantum.2.030335}.
	
	When slot $k$ is reached, its channel acts on $H_k^I\to H_k^O$ tensored with the identity on the private memory. It is $\mathcal A$ or $\mathcal B$ according to the fixed label of that slot. The composed map for one complete history is generally trace-nonincreasing. Summing the maps over all complete histories gives one deterministic channel
	$\mathcal C_{M,N}[\mathcal A,\mathcal B]: \Lop(P)\longrightarrow\Lop(F).$
	A fixed-order quantum comb is the special QC--CC case in which only one next-slot outcome is nonzero at each history \cite{PhysRevA.80.022339,PRXQuantum.2.030335}. When $M=N=0$, the simulator has no open slot and is simply a channel $\mathcal C_{0,0}:\Lop(P)\to\Lop(F)$ chosen independently of $\mathcal A$ and $\mathcal B$.
	
	\subsection{Exact universal simulation problem}
	\label{subsec:supp_exact_universal_simulation}
	
	Fix $d,M,N$ before the unknown channels are supplied. The exact simulation problem asks whether there exists one deterministic QC--CC simulator $\mathcal C_{M,N}$ such that
	\begin{equation}
		\forall\,\mathcal A,\mathcal B\in\CPTP_d,
		\qquad
		\mathcal C_{M,N}[\mathcal A,\mathcal B]
		=\mathsf S_d^{(2)}(\mathcal A,\mathcal B).
		\label{eq:supp_exact_two_label_task}
	\end{equation}

The same simulator must work for every CPTP pair. The counts are exact and the history sum is deterministic. Equality in Eq.~\eqref{eq:supp_exact_two_label_task} means equality of the complete channels from $P$ to $F$, including their extensions by the identity map on any reference. The resource domain is $\mathbb Z_{\geq0}^2$, including $M=0$ and $N=0$.

\begin{figure}[t]
		\centering
		\includegraphics[width=\linewidth]{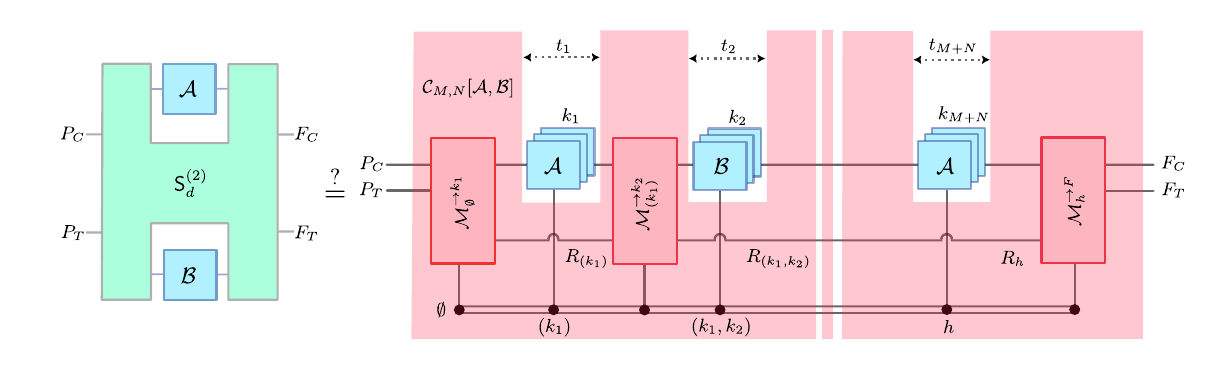}
		\caption{\textbf{Universal QC--CC simulation problem for the
		two-channel quantum switch.}
		The process on the left is the quantum switch
		$\mathsf S_d^{(2)}$, which uses one call to each unknown channel
		and retains coherence between the orders
		$\mathcal B\circ\mathcal A$ and $\mathcal A\circ\mathcal B$.
		The shaded region on the right represents a deterministic QC--CC
		in the circuit form of Ref.~\cite{PRXQuantum.2.030335}. It contains
		$M$ slots labelled by $\mathcal A$ and $N$ slots labelled by
		$\mathcal B$. All slots with the same label are filled by the same
		memoryless channel. At time $t_n$, the classical history determines
		one unused slot $k_n$. The layered channel boxes show the available
		choices, and the front box is the call made along the displayed
		history. The sequence $\mathcal A,\mathcal B,\ldots,\mathcal A$ is
		one representative complete history among those summed in
		$\mathcal C_{M,N}[\mathcal A,\mathcal B]$. The single lower wire
		carries the private quantum memory, with successive spaces
		$R_{(k_1)},R_{(k_1,k_2)},\ldots,R_h$, past the channel calls. The
		double wire records the classical history
		$\emptyset,(k_1),(k_1,k_2),\ldots,h$. The question mark asks whether,
		for fixed $M$ and $N$, one such simulator can satisfy
		$\mathcal C_{M,N}[\mathcal A,\mathcal B]
		=\mathsf S_d^{(2)}(\mathcal A,\mathcal B)$
		for every $\mathcal A,\mathcal B\in\CPTP_d$.}
		\label{fig:supp_switch_qccc}
	\end{figure}

	The quantum switch itself has a QC--QC realization with one call to each
	channel \cite{PRXQuantum.2.030335}. Figure~\ref{fig:supp_switch_qccc}
	contrasts its coherent control of the two channel orders with the
	classically recorded histories that define the QC--CC in
	Eq.~\eqref{eq:supp_exact_two_label_task}. Thus $(M,N)=(1,1)$ is the
	native comparison point for labelled queries when quantum control of
	causal order is allowed.
	
For channels of dimension $d=2^q$, the following result gives a lower bound on queries to one input when the other is queried once, under a mixed-unitary and unitary promise.

	\begin{predecessortheorem}[Ref.~\cite{Bavaresco2025}, Theorem~2]
		\label{thm:supp_predecessor_one_sided}
		Let $d=2^q$, where $q$ is a positive integer. If
		$1\leq M\leq\max\{2,d-1\}$, no deterministic QC--CC simulator
		$\mathcal C_{M,1}$ satisfies Eq.~\eqref{eq:supp_exact_two_label_task}
		for every mixed-unitary channel $\mathcal A$ and every unitary channel
		$\mathcal B$.
	\end{predecessortheorem}

	The exchange $\mathcal A\leftrightarrow\mathcal B$, accompanied by fixed Pauli-$X$ conjugations on the input and output control, maps the switched channel back to the convention in Eq.~\eqref{eq:supp_two_switch_kraus}. These fixed pre- and post-processing channels preserve the QC--CC class and the exact call counts. The theorem therefore also excludes $\mathcal C_{1,N}$ for
	$1\leq N\leq\max\{2,d-1\}$ when $\mathcal A$ is unitary and
	$\mathcal B$ is mixed unitary. The two statements give a one-sided lower bound of order $2^q$ on the number of labelled queries. Those bounds left a region in which both positive query counts scale with $d$ unresolved. The same source separately excludes selected finite allocations, including $(M,N)=(2,2)$, by a computer-assisted analysis \cite{Bavaresco2025}.

	The broader deterministic problem was left as an explicit conjecture.
	\begin{predecessorconjecture}[Ref.~\cite{Bavaresco2025}, Conjecture~1]
		\label{conj:supp_predecessor_two_sided}
		For $d=2^q$, there exists a function $g(q)=\Theta(2^q)$ such that no deterministic QC--CC simulator satisfies Eq.~\eqref{eq:supp_exact_two_label_task} for all $d$-dimensional channels whenever
		$\max\{M,N\}\leq g(q)$.
	\end{predecessorconjecture}

The impossibility proved below holds for every fixed finite pair of query counts and every target dimension $d\geq2$, establishing the conjectured exclusion and extending it beyond the proposed dimension-dependent range. Its universal CPTP hypothesis differs from the restricted mixed-unitary and unitary promise of Theorem~\ref{thm:supp_predecessor_one_sided}.

Section~\ref{sec:supp_finite_binary} first treats qubit channels with positive finite query counts $M,N$. Retaining selected calls and forming a uniform mixture with private classical randomness will connect the universal simulation promise to a circuit that queries each input once. Its fixed process tensors cannot satisfy the two causal normalizations. A physical encoding transfers this obstruction to every target dimension. Section~\ref{sec:supp_complete_r} applies the same retention construction to two selected inputs of the complete multichannel switch.

\section{Finite-query obstruction for the complete two-channel switch}
\label{sec:supp_finite_binary}

We first take $d=2$ and positive finite counts $M,N$, and suppose that the circuit of S1 reproduces the complete SWITCH for every pair of qubit channels. Replacing all but one occurrence of each input by identity channels gives a physical circuit with two calls. Its output, however, is not specified by the simulation promise, which concerns fillings by the same unknown channel at every occurrence of each label. We will identify the output of a uniform average of these circuits by applying that promise to mixtures of the identity and the unknown channels. To construct the circuits being averaged, we first show how to absorb known identity calls into the intervening operations. Allowing a general retained set and fixed preprocessing and postprocessing channels will also give the simulator reduction needed in S3.

\subsection{Retaining calls of a physical QC--CC circuit}
\begin{proposition}\label{prop:extraction}
Let $\mathcal C$ be a deterministic QC--CC circuit with a finite named inventory $\Omega$, and let $\emptyset\ne\Xi\subseteq\Omega$. Fill the removed slots with their identity channels, and fix channel-independent CPTP preprocessing and postprocessing channels $\mathcal E:\Lop(P')\to\Lop(P)$ and $\mathcal D:\Lop(F)\to\Lop(F')$. The resulting complete channel is realized by a deterministic QC--CC circuit with inventory exactly $\Xi$. This equality holds for every independent filling of the retained slots and every reference extension. Each named occurrence in $\Xi$ is used once, so the number of calls to a channel is the number of its retained occurrences.

\end{proposition}
\begin{proof}
Choose a physical realization of the named-slot circuit, before imposing any homogeneous-input promise. Deleting an identity call does not erase the classical information obtained before the next retained call. Two original histories with the same retained subsequence can leave different quantum memories and different sets of unused names. The reduced circuit must therefore retain the original history as well as the retained one.

For an original history $h=(k_1,\ldots,k_n)$, let $R_h$ be its finite memory, including the accessible classical records on which subsequent instruments depend. For a nonempty retained history $\gamma=(\gamma_1,\ldots,\gamma_m)$, let $\mathcal H_\gamma$ contain the original prefixes ending at $\gamma_m$ with retained subsequence $\gamma$. We store their memories in $\widehat R_\gamma\coloneqq\bigoplus_{h\in\mathcal H_\gamma}R_h$. The summand label records the classical history, while the state within that summand retains the quantum memory. The canonical embeddings $\iota_h:R_h\to\widehat R_\gamma$ satisfy $\iota_h^\dagger\iota_q=0$ for $h\ne q$, $\iota_h^\dagger\iota_h=\id_{R_h}$, and $\sum_h\iota_h\iota_h^\dagger=\id_{\widehat R_\gamma}$. If $h$ ends at $z$, their extensions to the slot input and output are $\iota_h^I\coloneqq\id_{H_z^I}\otimes\iota_h$ and $\iota_h^O\coloneqq\id_{H_z^O}\otimes\iota_h$.

At a nonterminal prefix, the CP outcome map $\mathcal M_h^{\to k}$ sends $\Lop(H_{k_n}^O\otimes R_h)$ to $\Lop(H_k^I\otimes R_{(h,k)})$ for each unused name $k\in\Omega\setminus\operatorname{used}(h)$. At the root the input is $P$, and at a complete prefix the terminal map has output $F$. These physical instruments can be taken on full input domains. In an effective-space description \cite[Sec.~IV and Appendix~C.2]{PRXQuantum.2.030335}, let $\Pi_h$ project onto the effective input of this named-slot realization. For one unused name $k_0$ and a density operator $\sigma_h$ in its output space, the extension
\begin{align}\label{eq:instrument-completion}
 \widetilde{\mathcal M}_h^{\to k}(X)
 \coloneqq\mathcal M_h^{\to k}(\Pi_h X\Pi_h)
       +\delta_{k,k_0}\Tr[(\id-\Pi_h)X]\sigma_h,\qquad
 \sum_k\widetilde{\mathcal M}_h^{\to k*}(\id)
 =\Pi_h\Bigl(\sum_k\mathcal M_h^{\to k*}(\id)\Bigr)\Pi_h+\id-\Pi_h=\id
\end{align}
is CP and preserves the instrument on its effective input, also with a reference. The effective input here is that of the process with independently filled named slots. At a complete prefix the same extension uses a replacement state on $F$. This extension is independent of the unknown channels. In the following construction we write $\mathcal M$ for these instruments on their full input spaces. At each interior prefix their Kraus operators obey
\begin{equation}\label{eq:original-instruments}
 \sum_{k\notin\operatorname{used}(h)}\sum_a
 M_{h,a}^{\to k\dagger}M_{h,a}^{\to k}
 =\id_{H_{k_n}^O R_h},\qquad 0<|h|<|\Omega|.
\end{equation}
The same completeness relation holds at the root and at a terminal history, with the corresponding domains,
\begin{equation}\label{eq:supp_retention_boundary_norm}
\sum_{k,a}M_{\emptyset,a}^{\to k\dagger}M_{\emptyset,a}^{\to k}=\id_P,
\qquad
\sum_aM_{h,a}^{\to F\dagger}M_{h,a}^{\to F}=\id_{H_{k_n}^O R_h}.
\end{equation}
At an interior history, normalization includes the sum over the next choice wherever that choice is still available.

Between the output of a retained call and the input of the next, all removed calls are known operations. Suppose the next retained name $z$ is reached through removed names $u_1,\ldots,u_s$. With $h_t=(h,u_1,\ldots,u_t)$ and $q'=(h_s,z)$, their chronological composition is
\begin{align}\label{eq:expanded-path}
 \mathcal T_{h\to q'}
 &\coloneqq\mathcal M_{h_s}^{\to z}\circ(\mathcal{I}_{u_s}\otimes\operatorname{id}_{R_{h_s}})
   \circ\mathcal M_{h_{s-1}}^{\to u_s}\circ\cdots\circ
   (\mathcal{I}_{u_1}\otimes\operatorname{id}_{R_{h_1}})\circ\mathcal M_h^{\to u_1},\notag\\
 T_{h\to q';\boldsymbol a}
 &=M_{h_s,a_s}^{\to z}(\id_{H_{u_s}}\otimes \id_{R_{h_s}})
   M_{h_{s-1},a_{s-1}}^{\to u_s}\cdots
   (\id_{H_{u_1}}\otimes \id_{R_{h_1}})M_{h,a_0}^{\to u_1}.
\end{align}
Here $\mathcal{I}_u$ is the identity channel at the removed slot. The map acts from $\Lop(H_{k_n}^O\otimes R_h)$ to $\Lop(H_z^I\otimes R_{q'})$. When $h=\emptyset$ it starts on $P$, and when $s=0$ it is $\mathcal M_h^{\to z}$. For a complete continuation $q=(h,u_1,\ldots,u_s)$ after the last retained name, replace the final map in this product by $\mathcal M_q^{\to F}$. Denote this fixed tail by $\mathcal T_{h\to q,F}$, and its sum over complete extensions by $\mathcal T_{h\to F}\coloneqq\sum_{q\supseteq h}\mathcal T_{h\to q,F}$. If no removed call remains, this continuation is the original terminal channel.

Let $\mathcal H_{\gamma z}(h)$ be the extensions of $h$ whose next retained name is $z$. For $X\in\Lop(H_{\gamma_m}^O\otimes\widehat R_\gamma)$, its $h$th diagonal block is $X_{hh}\coloneqq\iota_h^{O\dagger}X\iota_h^O$. The reduced instrument applies each path map to this block and embeds its output in the summand labelled by $q'$. The Kraus expansion gives
\begin{gather}
 \vcenter{\hbox{\includegraphics{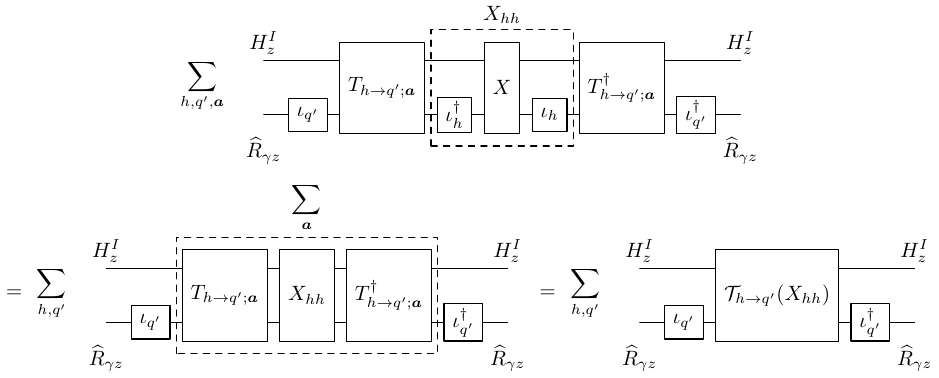}}}\notag\\[5pt]
 \begin{aligned}
 \widehat{\mathcal M}_\gamma^{\to z}(X)
 &\coloneqq\sum_{h\in\mathcal H_\gamma}\sum_{q'\in\mathcal H_{\gamma z}(h)}\sum_{\boldsymbol a}
 \iota_{q'}^I T_{h\to q';\boldsymbol a}
       (\iota_h^{O\dagger}X\iota_h^O)T_{h\to q';\boldsymbol a}^{\dagger}\iota_{q'}^{I\dagger}\\
 &=\sum_{h\in\mathcal H_\gamma}\sum_{q'\in\mathcal H_{\gamma z}(h)}
 \iota_{q'}^I\mathcal T_{h\to q'}(X_{hh})\iota_{q'}^{I\dagger}.
 \end{aligned}\label{eq:reduced-maps}
\end{gather}
The output belongs to $\Lop(H_z^I\otimes\widehat R_{\gamma z})$. At the root the same path construction starts on $P$. After the complete retained history $\gamma_{\rm fin}$, it starts on $H_{\gamma_{\rm last}}^O\otimes\widehat R_{\gamma_{\rm fin}}$ and sums the remaining tails to $F$. The initial instrument and terminal channel are therefore
\begin{align}\label{eq:reduced-boundaries}
 \widehat{\mathcal M}_\emptyset^{\to z}(X)
 \coloneqq\sum_{q'\in\mathcal H_z}\iota_{q'}^I
        \mathcal T_{\emptyset\to q'}(X)\iota_{q'}^{I\dagger},\qquad
 \widehat{\mathcal M}_{\gamma_{\rm fin}}^{\to F}(X)
 \coloneqq\sum_{h\in\mathcal H_{\gamma_{\rm fin}}}
       \mathcal T_{h\to F}(\iota_h^{O\dagger}X\iota_h^O),
\end{align}
with outputs on $H_z^I\otimes\widehat R_z$ and $F$, respectively. Compression, physical composition and orthogonal embedding make all these maps CP on their entire declared input spaces.

The normalization of the reduced instrument follows by summing all paths up to the next retained call. Fix an arbitrary reference $R$. In the following calculations, $\Tr_{\rm loc}$ traces over all systems except $R$, on the input or output space of the map to which it is applied. Every removed call uses the single identity Kraus operator $\id_{H_u}$ and preserves the reference marginal, since $\Tr_{\rm loc}[(\mathcal{I}_u\otimes\operatorname{id})(Z)]=\Tr_{\rm loc}[(\id_{H_u}\otimes \id)Z(\id_{H_u}\otimes \id)]=\Tr_{\rm loc}Z$.
At prefix $h$ with retained subsequence $\gamma$, the remaining retained and removed sets are $\Xi_\gamma\coloneqq\Xi\setminus\operatorname{used}(\gamma)$ and $\mathcal U_h\coloneqq(\Omega\setminus\Xi)\setminus\operatorname{used}(h)$. Every endpoint in $\mathcal H_{\gamma z}(h)$ either follows $h$ directly by $z$, or first follows it by a unique $u\in\mathcal U_h$. This gives the disjoint partition $\mathcal H_{\gamma z}(h)=\{(h,z)\}\mathbin{\dot\cup}\mathop{\dot\bigcup}_{u\in\mathcal U_h}\mathcal H_{\gamma z}(h,u)$.
For $\Xi_\gamma\ne\emptyset$, the path effect splits into the direct contributions and the continuations through the first removed slot. Tensoring this effect identity with $\id_R$ and contracting it with $Z$ gives the corresponding partial-trace recursion, by cyclicity on the circuit systems. For a first step through $u$, the intermediate operator is $Z_u\coloneqq(\mathcal{I}_u\otimes\operatorname{id})(\mathcal M_h^{\to u}\otimes\operatorname{id}_{\Lop(R)})(Z)$.

When $\mathcal U_h$ is empty, the endpoint partition contains only the direct branches and normalization is instrument completeness. Otherwise each $Z_u$ has one fewer removed slot available. The root uses this same partition with the initial instrument on $P\otimes R$. Induction on $|\mathcal U_h|$ gives
\begingroup
\predisplaypenalty=0
\begin{gather}
 \vcenter{\hbox{\includegraphics{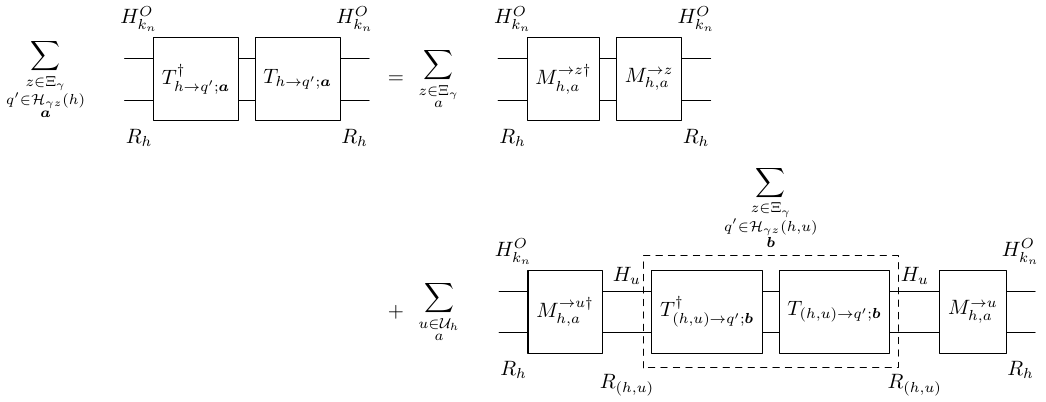}}}\notag\\[5pt]
 \begin{aligned}
 &\sum_{z\in\Xi_\gamma}\sum_{q'\in\mathcal H_{\gamma z}(h)}
    \Tr_{\rm loc}(\mathcal T_{h\to q'}\otimes\operatorname{id}_{\Lop(R)})(Z)\\
 &=\sum_{z\in\Xi_\gamma}\Tr_{\rm loc}(\mathcal M_h^{\to z}\otimes\operatorname{id}_{\Lop(R)})(Z)
   +\sum_{u\in\mathcal U_h}\sum_{z\in\Xi_\gamma}
       \sum_{q'\in\mathcal H_{\gamma z}(h,u)}
          \Tr_{\rm loc}(\mathcal T_{(h,u)\to q'}\otimes\operatorname{id}_{\Lop(R)})(Z_u)\\
 &=\sum_{z\in\Xi_\gamma}\Tr_{\rm loc}(\mathcal M_h^{\to z}\otimes\operatorname{id}_{\Lop(R)})(Z)
    +\sum_{u\in\mathcal U_h}\Tr_{\rm loc}Z_u\\
 &=\sum_{k\notin\operatorname{used}(h)}
       \Tr_{\rm loc}(\mathcal M_h^{\to k}\otimes\operatorname{id}_{\Lop(R)})(Z)
 =\Tr_{\rm loc}Z.
 \end{aligned}\label{eq:path-normalization}
\end{gather}
\endgroup
After every retained call has occurred, the unused names are precisely $\mathcal U_h$. The terminal path sum instead ends at the original channel to $F$. When no removed name remains, the terminal channel preserves the reference marginal. Otherwise the partition by the first removed call gives
\begin{align}\label{eq:terminal-path-normalization}
 \Tr_{\rm loc}(\mathcal T_{h\to F}\otimes\operatorname{id}_{\Lop(R)})(Z)
 &=\sum_{u\in\mathcal U_h}\Tr_{\rm loc}
      (\mathcal T_{(h,u)\to F}\otimes\operatorname{id}_{\Lop(R)})(Z_u)\notag\\
 &=\sum_{u\in\mathcal U_h}\Tr_{\rm loc}
      (\mathcal M_h^{\to u}\otimes\operatorname{id}_{\Lop(R)})(Z)=\Tr_{\rm loc}Z
 \qquad(\mathcal U_h\ne\emptyset).
\end{align}
Substituting these path sums into the reduced maps proves normalization even on inputs with off-diagonal prefix blocks. For a reference-extended input, the diagonal blocks are $X_{hh}=(\iota_h^{O\dagger}\otimes \id_R)X(\iota_h^O\otimes \id_R)$, and
\begin{align}\label{eq:reduced-normalization}
 \sum_{z\in\Xi_\gamma}\Tr_{\rm loc}
       (\widehat{\mathcal M}_\gamma^{\to z}\otimes\operatorname{id}_{\Lop(R)})(X)
 &=\sum_{z,h}\sum_{q'\in\mathcal H_{\gamma z}(h)}
       \Tr_{\rm loc}(\mathcal T_{h\to q'}\otimes\operatorname{id}_{\Lop(R)})(X_{hh})\notag\\
 &=\sum_h\Tr_{\rm loc}X_{hh}
 =\Tr_{\rm loc}\left[X\left(\id\otimes\sum_h\iota_h\iota_h^\dagger\otimes \id_R\right)\right]
 =\Tr_{\rm loc}X.
\end{align}
\newpage 
The root is normalized by Eq.~\eqref{eq:path-normalization}, and the terminal map by Eq.~\eqref{eq:terminal-path-normalization} and the same sum over $h$. Off-diagonal prefix blocks have zero partial trace over the record. The reduced circuit sums the classical alternatives while preserving quantum coherence within each memory sector.

To identify the output channel, take one complete original history $q=(k_1,\ldots,k_n)$ and write $q_l=(k_1,\ldots,k_l)$. Its retained calls occur at positions $r_1<\cdots<r_m$, with $z_j=k_{r_j}$ and $h_j=q_{r_j}$. Cutting the chronological circuit immediately before and after these calls leaves the initial segment from $\emptyset$ to $h_1$, the segments from $h_j$ to $h_{j+1}$, and the terminal tail from $h_m$ to $q,F$, all defined in Eq.~\eqref{eq:expanded-path}.

Let $a_0,\ldots,a_n$ index the original internal Kraus operators, and collect their consecutive segment indices into $\boldsymbol a_0,\ldots,\boldsymbol a_m$. If $O_{z,b}:H_z^I\to H_z^O$ is a Kraus operator of an independently chosen retained filling, define $Q_{k_{r_j},b_{r_j}}=O_{z_j,b_{r_j}}$ and use the single identity Kraus operator at each removed position. Write $\gamma_j=(z_1,\ldots,z_j)$. At a retained call, the embeddings obey $\iota_h^{O\dagger}(O_{z,b}\otimes\id_{\widehat R_\gamma})\iota_{h'}^I=O_{z,b}\otimes\iota_h^\dagger\iota_{h'}$, so only matching original prefixes contribute when the reduced instruments are composed. Their Kraus product reconstructs the original one,
\begin{align}\label{eq:kraus-reassembly}
 \widehat L
 &=T_{h_m\to q,F;\boldsymbol a_m}\iota_{h_m}^{O\dagger}
    (O_{z_m,b_{r_m}}\otimes \id_{\widehat R_{\gamma_m}})
    \iota_{h_m}^I T_{h_{m-1}\to h_m;\boldsymbol a_{m-1}}\cdots\iota_{h_1}^{O\dagger}
    (O_{z_1,b_{r_1}}\otimes \id_{\widehat R_{\gamma_1}})
    \iota_{h_1}^I T_{\emptyset\to h_1;\boldsymbol a_0}\notag\\
 &=T_{h_m\to q,F;\boldsymbol a_m}(O_{z_m,b_{r_m}}\otimes \id_{R_{h_m}})
    T_{h_{m-1}\to h_m;\boldsymbol a_{m-1}}\cdots
    (O_{z_1,b_{r_1}}\otimes \id_{R_{h_1}})T_{\emptyset\to h_1;\boldsymbol a_0}\notag\\
 &=M_{q,a_n}^{\to F}(Q_{k_n,b_n}\otimes \id_{R_q})
    M_{q_{n-1},a_{n-1}}^{\to k_n}\cdots
    M_{q_1,a_1}^{\to k_2}(Q_{k_1,b_1}\otimes \id_{R_{q_1}})
    M_{\emptyset,a_0}^{\to k_1}
 \coloneqq L_{q,\boldsymbol a,\boldsymbol b}.
\end{align}
Equation~\eqref{eq:expanded-path} inserts every removed call and internal map at its original position. For $m=1$ there are no intermediate segment factors. Cutting an original history at its retained calls gives exactly one sequence of the path operators in Eq.~\eqref{eq:kraus-reassembly}. Conversely, every nonzero reduced Kraus product concatenates to exactly one original history.

Let $E_\epsilon,D_\delta$ be Kraus operators of the fixed preprocessing and postprocessing channels. Reindexing the Kraus outer products using this correspondence gives, for every operator $X$ on $P'\otimes R$,
\begin{align}\label{eq:complete-retention}
 (\widehat{\mathcal C}_\Xi[\mathcal O]\otimes\operatorname{id}_{\Lop(R)})(X)
 &=\sum_{\delta,\epsilon}\sum_{\text{reduced products}}
   (D_\delta\widehat L E_\epsilon\otimes \id_R)X
   (D_\delta\widehat L E_\epsilon\otimes \id_R)^\dagger\notag\\
 &=\sum_{\delta,\epsilon}\sum_{q,\boldsymbol a,\boldsymbol b}
   (D_\delta L_{q,\boldsymbol a,\boldsymbol b}E_\epsilon\otimes \id_R)X
   (D_\delta L_{q,\boldsymbol a,\boldsymbol b}E_\epsilon\otimes \id_R)^\dagger\notag\\
 &=(\mathcal D\otimes\operatorname{id}_{\Lop(R)})
   (\mathcal C[\mathcal O_\Xi,\mathcal{I}_{\Omega\setminus\Xi}]\otimes\operatorname{id}_{\Lop(R)})
   (\mathcal E\otimes\operatorname{id}_{\Lop(R)})(X).
\end{align}
The preprocessing channel composes with the initial instrument, and the postprocessing channel with the terminal channel. Their summed effects are $\mathcal E^*(\id_P)=\id_{P'}$ and $(\widehat{\mathcal M}_{\gamma_{\rm fin}}^{\to F})^*(\mathcal D^*(\id_{F'}))=\id$, respectively. They add no calls. Thus the reduced QC--CC circuit uses exactly the retained occurrences and realizes the stated channel, including its action on the reference.

\end{proof}

\subsection{The complete channel of the selected-pair average}
\begin{corollary}\label{cor:positive-average}
Suppose a deterministic QC--CC circuit satisfies Eq.~\eqref{eq:supp_exact_two_label_task} for all qubit channels, with $M,N\geq1$. The uniform mixture obtained by retaining one of its $M$ calls to $\mathcal A$ and one of its $N$ calls to $\mathcal B$, and filling the other calls with identities, uses each unknown channel once and realizes
$\mathsf S_2^{(2)}((1-1/M)\mathcal I+\mathcal A/M,(1-1/N)\mathcal I+\mathcal B/N)$.
Here $\mathcal I$ is the qubit identity channel. The equality holds for the complete channel and every reference extension.
\end{corollary}
\begin{proof}
For each $i\in[M]$ and $j\in[N]$, apply Proposition~\ref{prop:extraction} with retained set $\{A_i,B_j\}$ and identity preprocessing and postprocessing. Denote the resulting complete channel by $\mathcal C_{\{i\},\{j\}}[\mathcal A,\mathcal B]$, with the reduced instruments and memory spaces constructed for this retained pair. Sampling the pair uniformly with private classical randomness gives the average $\overline{\mathcal C}[\mathcal A,\mathcal B]\coloneqq(MN)^{-1}\sum_{i,j}\mathcal C_{\{i\},\{j\}}[\mathcal A,\mathcal B]$. 

\noindent\begin{minipage}{\linewidth}
\noindent Resolving each selected circuit into its two chronological orders gives the contraction below, with an arbitrary reference $R$ carried unchanged,

\begin{gather}
 \vcenter{\hbox{\includegraphics{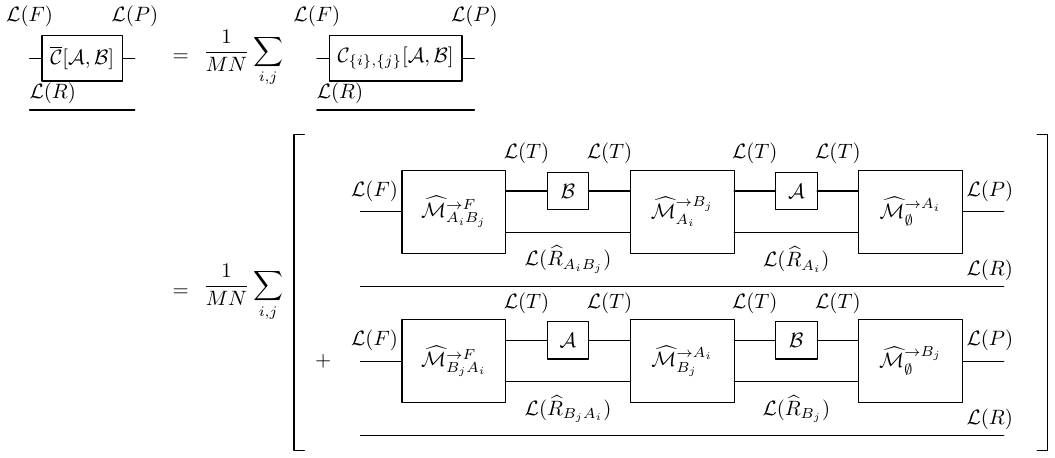}}}\notag\\[5pt]
\begin{aligned}
 \mathcal C_{\{i\},\{j\}}[\mathcal A,\mathcal B]
 &=\widehat{\mathcal M}_{A_iB_j}^{\to F}\circ(\mathcal B\otimes\operatorname{id})
   \circ\widehat{\mathcal M}_{A_i}^{\to B_j}\circ(\mathcal A\otimes\operatorname{id})
    \circ\widehat{\mathcal M}_\emptyset^{\to A_i}\\
 &\quad+\widehat{\mathcal M}_{B_jA_i}^{\to F}\circ(\mathcal A\otimes\operatorname{id})
   \circ\widehat{\mathcal M}_{B_j}^{\to A_i}\circ(\mathcal B\otimes\operatorname{id})
   \circ\widehat{\mathcal M}_\emptyset^{\to B_j}.
\end{aligned}\label{eq:two-call-channel}
\end{gather}
\end{minipage}\par
The identity maps act on the memories carried by the corresponding instruments. The averaged circuit is a deterministic QC--CC circuit using each unknown channel once. For fixed input channels, its output Choi operator is positive and satisfies $\Tr_FJ^{\overline{\mathcal C}[\mathcal A,\mathcal B]}=\id_P$.

The initial effects for its two orders are the respective averages of $(\widehat{\mathcal M}_\emptyset^{\to A_i})^*(\id)$ and $(\widehat{\mathcal M}_\emptyset^{\to B_j})^*(\id)$. They sum to $\id_P$ and need not be scalar multiples of the identity, so the permitted dependence of the order choice on the input is retained.

It remains to identify the complete channel implemented by this average. Fix arbitrary qubit channels $\mathcal A,\mathcal B$ and vary the physical mixtures $\mathcal A_t\coloneqq(\mathcal{I}+t\mathcal A)/(1+t)$ and $\mathcal B_s\coloneqq(\mathcal{I}+s\mathcal B)/(1+s)$, with $t,s\geq0$. Write $\mathcal C_{S,T}[\mathcal A,\mathcal B]$ for the original circuit with $\mathcal A$ in the occurrences indexed by $S\subseteq[M]$, $\mathcal B$ in those indexed by $T\subseteq[N]$, and identity channels elsewhere. The simulation promise fixes the output when all occurrences of each label contain the same mixture. To determine the average, we compare that homogeneous filling with the expansion over independently named slots.

The corresponding target polynomial can be computed on the full control and reference interface. If $\{A_\alpha\}_\alpha$ and $\{B_\beta\}_\beta$ are Kraus families of the fixed channels, those of the mixtures are $\{\id/\sqrt{1+t}\}\cup\{\sqrt{t/(1+t)}A_\alpha\}_\alpha$ and $\{\id/\sqrt{1+s}\}\cup\{\sqrt{s/(1+s)}B_\beta\}_\beta$. Write $K_{0,\alpha\beta}=B_\beta A_\alpha$ and $K_{1,\alpha\beta}=A_\alpha B_\beta$. Substitution into Eq.~\eqref{eq:supp_two_switch_kraus}, using the same Kraus label in both control sectors, gives
\begin{align}\label{eq:diluted-blocks}
 &(1+t)(1+s)(\SW_2^{(2)}(\mathcal A_t,\mathcal B_s)\otimes\operatorname{id}_{\Lop(R)})
       (\ket c\bra{c'}\otimes X_{TR})=\ket c\bra{c'}\otimes\Bigl[X_{TR}
   +t\sum_\alpha(A_\alpha\otimes \id_R)X_{TR}(A_\alpha^\dagger\otimes \id_R)\notag\\
 &\hspace{4em}+s\sum_\beta(B_\beta\otimes \id_R)X_{TR}(B_\beta^\dagger\otimes \id_R)+ts\sum_{\alpha,\beta}(K_{c,\alpha\beta}\otimes \id_R)X_{TR}
                  (K_{c',\alpha\beta}^\dagger\otimes \id_R)\Bigr]
\end{align}
for every $c,c'\in\{0,1\}$ and every target and reference operator $X_{TR}$. The single-channel terms coincide in the two orders. The last term retains their coherent control, including its off-diagonal blocks. Thus multilinearity of the actual circuit, its universal simulation property, and the complete target calculation give successive equalities for the same output polynomial,
\begin{align}
 &\sum_{S\subseteq[M]}\sum_{T\subseteq[N]}
       t^{|S|}s^{|T|}J^{\mathcal C_{S,T}[\mathcal A,\mathcal B]}=(1+t)^M(1+s)^N J^{\mathcal C_{M,N}[\mathcal A_t,\mathcal B_s]}
 =(1+t)^M(1+s)^N J^{\SW_2^{(2)}(\mathcal A_t,\mathcal B_s)}\notag\\
 &\quad=(1+t)^{M-1}(1+s)^{N-1}
   \Bigl[J^{\operatorname{id}_{\Lop(P_C\otimes P_T)}}+tJ^{\operatorname{id}_{\Lop(P_C)}\otimes\mathcal A}
       +sJ^{\operatorname{id}_{\Lop(P_C)}\otimes\mathcal B}
       +tsJ^{\SW_2^{(2)}(\mathcal A,\mathcal B)}\Bigr].
       \label{eq:complete-mixture-identity}
\end{align}
These finite operator-valued polynomials agree for all $t,s>0$, so their matrix entries agree coefficient by coefficient. The coefficient of $ts$ in the final row receives weights $(M-1)(N-1)$, $N-1$, $M-1$, and $1$ from its four terms. It follows that the physical average constructed above satisfies
\begin{align}\label{eq:positive-selected}
 J^{\overline{\mathcal C}[\mathcal A,\mathcal B]}
 &=\frac1{MN}\sum_{i=1}^M\sum_{j=1}^N
       J^{\mathcal C_{\{i\},\{j\}}[\mathcal A,\mathcal B]}\notag\\
 &=\frac{(M-1)(N-1)}{MN}J^{\operatorname{id}_{\Lop(P_C\otimes P_T)}}
   +\frac{N-1}{MN}J^{\operatorname{id}_{\Lop(P_C)}\otimes\mathcal A}
   +\frac{M-1}{MN}J^{\operatorname{id}_{\Lop(P_C)}\otimes\mathcal B}+\frac1{MN}J^{\SW_2^{(2)}(\mathcal A,\mathcal B)}
 .
\end{align}
The four weights are nonnegative and sum to one, including the cases $M=1$ or $N=1$. This proves the stated complete-channel identity. In particular, the one-use circuit realizes a mixture containing the complete SWITCH with strictly positive weight $1/(MN)$.
\end{proof}

\subsection{Control sectors and fixed physical tensors}
The averaged channel belongs to the family
\begin{equation}\label{eq:diluted}
\mathcal R_{\lambda_A,\lambda_B}(\mathcal A,\mathcal B)
=\mathsf S_2^{(2)}\bigl((1-\lambda_A)\mathcal I+\lambda_A\mathcal A,
(1-\lambda_B)\mathcal I+\lambda_B\mathcal B\bigr),
\end{equation}
with $\lambda_A=1/M$ and $\lambda_B=1/N$. We will rule out a one-use QC--CC realization for every $0<\lambda_A,\lambda_B\leq1$. Fix such a pair and suppose that a deterministic QC--CC circuit uses each qubit channel once and implements $\mathcal R_{\lambda_A,\lambda_B}(\Ad_U,\Ad_V)$ for all $U,V\in\mathrm U(2)$, where $\Ad_U(X)=UXU^\dagger$. Any circuit realizing the average for all CPTP inputs satisfies this restriction.

Group the circuit histories according to whether $A$ or $B$ is queried first. Let $\mathcal P_A,\mathcal P_B$ be its initial CP outcome maps, with effects $E_o\coloneqq\mathcal P_o^*(\id)$ satisfying $E_A+E_B=\id_P$. The positive process operators $W_A,W_B$ include these outcomes. Choose rank-one decompositions independently of the oracle inputs,
\begin{equation}\label{eq:positive-process}
 W=W_A+W_B,\qquad
 W_o=\sum_\alpha\ketketbra{C_o^{(\alpha)}}{C_o^{(\alpha)}},\qquad o\in\{A,B\}.
\end{equation}
The vectors retain the past, both open slots and the future in the fixed order
\begin{equation}\label{eq:wire-order}
 \ketket{C_o^{(\alpha)}}\in P_C\otimes P_T\otimes H_A^I\otimes H_A^O
 \otimes H_B^I\otimes H_B^O\otimes F_C\otimes F_T,
\end{equation}
with all factors two-dimensional. Filling the two slots of a fixed vector produces an operator $G_{o\alpha}(X,Y):P\to F$ that depends bilinearly on $X,Y$. Let $p,f$ index the past and future control factors, and let $t,u$ index the corresponding target factors. With the input-first vectorization and bilinear vector link of S1,
\begin{equation}\label{eq:amplitude}
 [G_{o\alpha}(X,Y)]_{fu,pt}
 =\sum_{i,j,k,l}\bra{p,t,i,j,k,l,f,u}\ketket{C_o^{(\alpha)}}X_{ji}Y_{lk}.
\end{equation}

For unitary inputs the complete SWITCH is a unitary channel. Its operator extends bilinearly to
\begin{equation}\label{eq:w}
 w(X,Y)\coloneqq\ket0_{F_C}\bra0_{P_C}\otimes YX+\ket1_{F_C}\bra1_{P_C}\otimes XY.
\end{equation}
Write $J(U,V)\coloneqq J^{\mathcal R_{\lambda_A,\lambda_B}(\Ad_U,\Ad_V)}$ for the complete output Choi operator. Evaluating it from the circuit and from Eq.~\eqref{eq:diluted} gives
\begin{align}\label{eq:target-choi}
 \sum_{o,\alpha}\ketketbra{G_{o\alpha}(U,V)}{G_{o\alpha}(U,V)}
 &=\lambda_A\lambda_B\ketketbra{w(U,V)}{w(U,V)}+\lambda_A(1-\lambda_B)\ketketbra{\id_C\otimes U}{\id_C\otimes U}\notag\\
 &\quad+(1-\lambda_A)\lambda_B\ketketbra{\id_C\otimes V}{\id_C\otimes V}+(1-\lambda_A)(1-\lambda_B)\ketketbra{\id_C\otimes \id}{\id_C\otimes \id}.
\end{align}
For every vector $v$ orthogonal to the four displayed amplitudes, positivity gives
\begin{equation}\label{eq:positive-support}
 \bra v J(U,V)\ket v=\sum_{o,\alpha}\left|\bra v\ketket{G_{o\alpha}(U,V)}\right|^2=0.
\end{equation}
Every linked vector therefore lies in their span. This support statement also holds when $\lambda_A=1$ or $\lambda_B=1$. Its coefficients could initially depend on both matrices, and the spanning vectors become dependent for commuting inputs. The causal argument needs a decomposition of the fixed vector before either slot is filled, valid also at those dependent inputs. The difference of its control blocks will determine the coefficient of the SWITCH tensor. The common block must then account for the three remaining target paths, through the first slot, through the second slot, or directly from the past target to the future target.

The unfilled SWITCH tensor joins the same target wires through its two orders,
\begin{align}\label{eq:unfilled-switch}
 \ketket S={}&\ket0_{P_C}\ketket \id_{P_TH_A^I}
       \ketket \id_{H_A^OH_B^I}\ketket \id_{H_B^OF_T}\ket0_{F_C}+\ket1_{P_C}\ketket \id_{P_TH_B^I}
       \ketket \id_{H_B^OH_A^I}\ketket \id_{H_A^OF_T}\ket1_{F_C}.
\end{align}
The vector link of $\ketket S$ with $\ketket X\otimes\ketket Y$ is $\ketket{w(X,Y)}$. This identifies the coefficient of $w$ in the operator representation with the coefficient of $\ketket S$ in the fixed process vector.

\begin{lemma}\label{lem:physical}
Let $\ketket C$ be a fixed vector on the labelled wires of Eq.~\eqref{eq:wire-order}, and suppose its linked operator $G_C(U,V)$ belongs, for every unitary pair, to the span of
$w(U,V)$, $\id_C\otimes \id$, $\id_C\otimes U$ and $\id_C\otimes V$.
There are unique fixed $e\in\C$ and tensors $\zeta^{(b)}$, $\zeta^{(c)}$, $\zeta^{(0)}$
on the indicated slots such that
\begin{align}\label{eq:physical-vector}
 \ketket C={}&e\ketket S+\ketket \id_{P_CF_C}\otimes\Bigl[
 \ketket \id_{P_TH_A^I}\ketket \id_{H_A^OF_T}
                  \ketket{\zeta^{(b)}}_{H_B^IH_B^O}\notag\\
 &\hspace{2em}+\ketket \id_{P_TH_B^I}\ketket \id_{H_B^OF_T}
                  \ketket{\zeta^{(c)}}_{H_A^IH_A^O}+\ketket \id_{P_TF_T}
                  \ketket{\zeta^{(0)}}_{H_A^IH_A^OH_B^IH_B^O}\Bigr].
\end{align}

The fixed tensors determine linear forms $b,c$ and a bilinear scalar form $\ell_0$ by contraction with the slot inputs.
\end{lemma}
\begin{proof}
We establish this physical decomposition through the bilinear amplitude of the fixed vector.
Its required form is
\begin{equation}\label{eq:normal-form}
 G(X,Y)=e\,w(X,Y)+\id_C\otimes
       \bigl[\ell_0(X,Y)\id+b(Y)X+c(X)Y\bigr]
 \qquad(X,Y\in M_2(\C)).
\end{equation}
Write $G_{fp}\coloneqq(\bra f_{F_C}\otimes \id)G(\ket p_{P_C}\otimes \id)$. We first extend polynomial identities for these amplitudes from unitary inputs to complex matrices. If a polynomial $q$ vanishes on $\mathrm U(2)$, choose a real Hermitian basis $H_1,\ldots,H_4$ and consider $q(\exp(i\sum_j z_jH_j))$. It vanishes for real $z_j$. Successive applications of the one-variable identity theorem make it vanish for complex $z_j$ as well. The derivative of $z\mapsto\exp(i\sum_jz_jH_j)$ at the origin is $v\mapsto i\sum_jv_jH_j$, which is invertible because the Hermitian basis is also a complex basis of $M_2(\C)$. Its image therefore contains a complex neighborhood of the identity, on which  $q$ vanishes, and $q$ is the zero polynomial.

For two inputs, apply the argument first with the second input unitary and fixed, and then with the first input complex and fixed. This continuation concerns polynomial amplitude identities. The positive Choi equality in Eq.~\eqref{eq:target-choi} retains its physical unitary domain.

The support hypothesis makes $G_{01}=G_{10}=0$ on unitary pairs and hence on all matrix pairs. Set $B(X,Y)\coloneqq G_{11}(X,Y)-G_{00}(X,Y)$. On unitary inputs the terms proportional to $\id_C$ cancel, leaving $B(U,V)$ proportional to $[U,V]$. In particular, $B(\id,V)=B(U,\id)=B(U,U)=0$. Polynomial continuation gives the same identities on all matrices, as well as $\Tr B(X,Y)=0$ and $B(X,Y)\wedge[X,Y]=0$, where the exterior product in $\Lambda^2M_2(\C)$ records linear dependence of the two output matrices. Bilinearity now gives $B(X,Y)+B(Y,X)=B(X+Y,X+Y)-B(X,X)-B(Y,Y)=0$. 

Thus $B$ depends only on the traceless parts $x=X-\tfrac12\Tr(X)\id$ and $y=Y-\tfrac12\Tr(Y)\id$ and defines an alternating bilinear map on $\mathfrak g\coloneqq\mathfrak{sl}_2(\C)$.

The bracket map $\Lambda^2\mathfrak g\to\mathfrak g$ is a linear isomorphism. For the matrix units $E_{ij}\coloneqq\ket i\bra j$, the identities $[E_{01},E_{10}]=E_{00}-E_{11}$ and $[E_{00}-E_{11},E_{01}]=2E_{01}$, $[E_{00}-E_{11},E_{10}]=-2E_{10}$ send a basis to a basis. Every bivector in this three-dimensional space is decomposable. Indeed, for a nonzero $\eta$, the map $z\mapsto z\wedge\eta$ into $\Lambda^3\mathfrak g$ is nonzero and has a two-dimensional kernel. Choose a basis $u,v$ of that kernel and complete it by $w$. In $\eta=a u\wedge v+b u\wedge w+c v\wedge w$, the two kernel equations give $b=c=0$. Hence every element of $\mathfrak g$ is a single commutator.

There is consequently a unique linear map $L:\mathfrak g\to\mathfrak g$ with $B(x,y)=L([x,y])$. Since every element is a commutator, the continued support identity gives $Lz\in\C z$ for every $z$. This forces $L$ to be scalar. If independent vectors $z,w$ have respective multipliers $a_z,a_w$, linearity at $z+w$ gives $a_z z+a_w w=a_{z+w}(z+w)$ and thus $a_z=a_w=a_{z+w}$. Proportional vectors are covered by linearity. Consequently one fixed scalar $e$, independent of both slot inputs, determines the control difference,
\begin{equation}\label{eq:global-switch-coefficient}
 G_{11}(X,Y)-G_{00}(X,Y)=e[X,Y].
\end{equation}

Subtracting this SWITCH contribution leaves equal diagonal control blocks, which define the bilinear map $H$ by $G(X,Y)-e\,w(X,Y)=\id_C\otimes H(X,Y)$. At a noncommuting unitary pair, the control difference fixes the SWITCH coefficient in the support expansion to be this same $e$. Hence $H(U,V)\in\Span\{\id,U,V\}$. For its traceless part $H_0(X,Y)\coloneqq H(X,Y)-\tfrac12\Tr(H(X,Y))\id$, trace cyclicity gives $\Tr(H_0(U,V)[U,V])=0$. This identity also holds at commuting pairs because their commutator vanishes. Polynomial continuation yields
\begin{equation}\label{eq:residual-trace}
 \Tr\bigl(H_0(X,Y)[X,Y]\bigr)=0
 \qquad(X,Y\in M_2(\C)).
\end{equation}
To determine the remaining fixed forms, we first restrict the common block to traceless inputs. The trace form $\beta(x,y)\coloneqq\Tr(xy)$ on $\mathfrak g$ is symmetric, complex-bilinear and nondegenerate. To see nondegeneracy, a traceless matrix orthogonal to $\mathfrak g$ is also orthogonal to $\id$, so testing against all matrix units makes it zero. Cyclicity gives its invariance directly, $\beta([x,y],z)=\Tr(xyz-yxz)=\Tr(xyz-xzy)=\beta(x,[y,z])$. In particular, $\beta(x,[x,y])=\beta(y,[x,y])=0$.

For $x\in\mathfrak g$, write $\operatorname{ad}_x(q)\coloneqq[x,q]$. When $x\ne0$, the bracket isomorphism gives $[x,q]=0$ exactly when $x\wedge q=0$, so $\ker\operatorname{ad}_x=\C x$. Invariance and nondegeneracy also give $(\operatorname{im}\operatorname{ad}_x)^{\perp_\beta}=\ker\operatorname{ad}_x$, because $\beta(q,[x,z])=\beta([q,x],z)$ vanishes for all $z$ precisely when $[q,x]=0$.

If $x,y$ are independent, $[x,y]\ne0$, and its orthogonal complement is the two-dimensional space $\Span\{x,y\}$. Equation~\eqref{eq:residual-trace} therefore puts $H_0(x,y)$ in this span. For the diagonal values, substitute $y=x+tz$ into the same identity to obtain $0=t\beta(H_0(x,x),[x,z])+t^2\beta(H_0(x,z),[x,z])$. The coefficient of $t$ vanishes for every $z$, so $H_0(x,x)\in(\operatorname{im}\operatorname{ad}_x)^{\perp_\beta}=\C x$ for nonzero $x$. Bilinearity includes zero and proportional pairs. Thus $H_0(x,y)\in\Span\{x,y\}$ for all traceless inputs, including nonzero nilpotent inputs. This determines its symmetric and alternating parts separately.

The alternating part of $H_0$ factors through the bracket isomorphism, so there is a unique endomorphism $T$ with $\tfrac12(H_0(x,y)-H_0(y,x))=T([x,y])$. Equation~\eqref{eq:residual-trace}, in both input orders, gives $\beta(T[x,y],[x,y])=0$. Since every element is a commutator, $\beta(Tz,z)=0$ for every $z\in\mathfrak g$. Applying this equality to $z+w$ and subtracting its values at $z$ and $w$ gives $\beta(Tz,w)+\beta(z,Tw)=0$, so $T\in\mathfrak{so}(\mathfrak g,\beta)$.

The adjoint representation supplies all endomorphisms skew with respect to $\beta$. Invariance gives $\beta([z_0,z],w)+\beta(z,[z_0,w])=0$, showing that $\operatorname{ad}(\mathfrak g)\subseteq\mathfrak{so}(\mathfrak g,\beta)$. The kernel of the adjoint representation is contained in the centralizers of two independent nonzero elements, whose intersection is zero by the centralizer identity above. On the other hand, nondegeneracy identifies a skew endomorphism $T$ with the alternating form $(z,w)\mapsto\beta(Tz,w)$. Conversely, an alternating bilinear form determines each $Tz$ uniquely through this pairing, and the resulting map is linear and skew. Thus $\dim\mathfrak{so}(\mathfrak g,\beta)=\dim\Lambda^2\mathfrak g^*=3=\dim\operatorname{ad}(\mathfrak g)$, and $T=\operatorname{ad}_{z_0}$ for a unique $z_0\in\mathfrak g$.

The alternating residual is therefore the double commutator $[z_0,[x,y]]$. To recover its coefficients on the input matrices, polarize the Cayley--Hamilton identity $x^2=\tfrac12\Tr(x^2)\id$ for traceless $2\times2$ matrices, obtaining $xy+yx=\Tr(xy)\id$. Applying this identity to the products containing $z_0$ yields
\begin{align}\label{eq:double-commutator}
 \tfrac12(H_0(x,y)-H_0(y,x))
 &=z_0xy-z_0yx-xyz_0+yxz_0\notag\\
 &=-x(z_0y+yz_0)+y(z_0x+xz_0)
       +\Tr(z_0x)y-\Tr(z_0y)x\notag\\
 &=2\Tr(z_0x)y-2\Tr(z_0y)x.
\end{align}
The diagonal values determine the symmetric part. For $x\ne0$, define $\ell(x)$ by $H_0(x,x)=2\ell(x)x$, and set $\ell(0)=0$. For independent $x,y$, the range condition gives $\tfrac12(H_0(x,y)+H_0(y,x))=a x+b' y$. Bilinearity at $x+ty$ gives $2\ell(x+ty)(x+ty)=2(\ell(x)+ta)x+2t(b'+t\ell(y))y$. For $t\ne0$, comparison of the two coefficients gives $\ell(x)+ta=\ell(x+ty)=b'+t\ell(y)$. Varying $t$ yields $a=\ell(y)$ and $b'=\ell(x)$. At $t=1$ this proves additivity for independent pairs. On each line, $H_0(sx,sx)=s^2H_0(x,x)$ implies $\ell(sx)=s\ell(x)$, with zero included by definition. It also gives additivity for dependent pairs, so $\ell\in\mathfrak g^*$.

Combining the symmetric and alternating parts defines the two coefficient forms that will occur on the physical slots. On $\mathfrak g$, set $b(y)\coloneqq\ell(y)-2\Tr(z_0y)$ and $c(x)\coloneqq\ell(x)+2\Tr(z_0x)$, giving $H_0(x,y)=b(y)x+c(x)y$. Their extension to scalar input directions is fixed by the original support hypothesis. At $(\id,V)$, the SWITCH amplitude is $\id_C\otimes V$, so $H(\id,V)\in\Span\{\id,V\}$. Likewise $H(U,\id)\in\Span\{\id,U\}$, and $H_0(\id,\id)=0$. These restrictions give the polynomial identities $H_0(\id,V)\wedge(V-\tfrac12\Tr(V)\id)=0$ and $H_0(U,\id)\wedge(U-\tfrac12\Tr(U)\id)=0$. Continuing each in its remaining input makes $y\mapsto H_0(\id,y)$ and $x\mapsto H_0(x,\id)$ preserve every line in $\mathfrak g$. By the linearity argument used for the control difference, they are scalar maps.

\newpage 
Denote their scalar multipliers by $c(\id)$ and $b(\id)$, respectively, so that $H_0(\id,y)=c(\id)y$ and $H_0(x,\id)=b(\id)x$. Together with the forms already defined on $\mathfrak g$, these values determine unique linear extensions of $b,c$ to $M_2(\C)=\C \id\oplus\mathfrak g$. For the traceless parts $x,y$ of $X,Y$, respectively, bilinearity now gives
\begin{align}\label{eq:scalar-restoration}
 H_0(X,Y)
 =\tfrac12\Tr(X)c(\id)y+\tfrac12\Tr(Y)b(\id)x+b(y)x+c(x)y=b(Y)x+c(X)y.
\end{align}
Consequently $H(X,Y)-b(Y)X-c(X)Y$ is scalar. Its coefficient $\ell_0(X,Y)\coloneqq\tfrac12\Tr[H(X,Y)-b(Y)X-c(X)Y]$ is bilinear, proving Eq.~\eqref{eq:normal-form} on the full matrix space.

Equation~\eqref{eq:normal-form} determines the linked operator on all input matrices. We now recover the fixed vector on its labelled wires. The forward contraction in Eq.~\eqref{eq:amplitude} gives
\begin{align}\label{eq:amplitude-recovery}
 \bra{p,t,i,j,k,l,f,u}\ketket C
 =[G_C(E_{ji},E_{lk})]_{fu,pt},\qquad
 G_C(X,Y)=\sum_{i,j,k,l}X_{ji}Y_{lk}G_C(E_{ji},E_{lk}).
\end{align}
As in the multilinearity argument of Lemma~3 in Ref.~\cite{Bavaresco2025}, matrix-unit evaluation identifies every tensor component. In the present argument the forms in Eq.~\eqref{eq:normal-form} supply those components. Define $\ketket{\zeta^{(b)}}\coloneqq\sum_{k,l}b(E_{lk})\ket k_{H_B^I}\ket l_{H_B^O}$ and $\ketket{\zeta^{(c)}}\coloneqq\sum_{i,j}c(E_{ji})\ket i_{H_A^I}\ket j_{H_A^O}$. The bilinear scalar term is represented on both slots by
\begin{equation}\label{eq:coefficient-tensors}
 \ketket{\zeta^{(0)}}\coloneqq\sum_{i,j,k,l}\ell_0(E_{ji},E_{lk})
 \ket i_{H_A^I}\ket j_{H_A^O}
 \ket k_{H_B^I}\ket l_{H_B^O}.
\end{equation}
The definitions use the bilinear vector link of S1. Denote the right-hand side of Eq.~\eqref{eq:physical-vector} by $\ketket{\widehat C}$.
The three terms in brackets transmit the target through the $\mathcal A$ slot, through the $\mathcal B$ slot, or directly from $P_T$ to $F_T$, respectively. The slots not traversed by the target carry precisely the coefficient forms just determined. This is a decomposition of the process amplitude. Causal normalization is imposed on $W_A$ and $W_B$. Contracting the three paths and the two SWITCH paths verifies the identification on the original labelled wires,
\begin{align}\label{eq:wire-contractions}
 [G_{\widehat C}(X,Y)]_{fu,pt}
 &=\delta_{pf}\sum_{i,j,k,l}X_{ji}Y_{lk}
 \Bigl[e\delta_{p0}\delta_{ti}\delta_{jk}\delta_{lu}
       +e\delta_{p1}\delta_{tk}\delta_{li}\delta_{ju}+\zeta^{(b)}_{kl}\delta_{ti}\delta_{ju}
       +\zeta^{(c)}_{ij}\delta_{tk}\delta_{lu}
       +\zeta^{(0)}_{ijkl}\delta_{tu}\Bigr]\notag\\
 &=\delta_{pf}\Bigl[e\delta_{p0}\sum_jY_{uj}X_{jt}
       +e\delta_{p1}\sum_iX_{ui}Y_{it}
       +X_{ut}\sum_{k,l}\zeta^{(b)}_{kl}Y_{lk}+Y_{ut}\sum_{i,j}\zeta^{(c)}_{ij}X_{ji}
       +\delta_{tu}\sum_{i,j,k,l}\zeta^{(0)}_{ijkl}X_{ji}Y_{lk}\Bigr]\notag\\
 &=\delta_{pf}\bigl[e\delta_{p0}(YX)_{ut}+e\delta_{p1}(XY)_{ut}
       +b(Y)X_{ut}+c(X)Y_{ut}+\ell_0(X,Y)\delta_{tu}\bigr]\notag\\
 &=[G_C(X,Y)]_{fu,pt}.
\end{align}
Evaluation on matrix units in Eq.~\eqref{eq:amplitude-recovery} recovers every component of the unfilled vector, and therefore $\ketket{\widehat C}=\ketket C$.

Finally, subtract two such representations. Their control difference at a noncommuting pair fixes $\Delta e=0$, leaving $\Delta\ell_0(X,Y)\id+\Delta b(Y)X+\Delta c(X)Y=0$. For nonscalar $Y$, choose $X\notin\Span\{\id,Y\}$. Independence gives $\Delta b(Y)=0$. Since $Y+\id$ is also nonscalar, linearity gives $\Delta b(\id)=\Delta b(Y+\id)-\Delta b(Y)=0$, and hence $\Delta b=0$ on all matrices. Interchanging $X,Y$ proves $\Delta c=0$, after which $\Delta\ell_0=0$. The coefficient tensors are therefore unique as well. The same fixed coefficients can now be used in the target identity and the causal constraints on each ordered process.
\end{proof}

\subsection{Target identities and causal normalization}
Lemma~\ref{lem:physical} gives fixed $e_{o\alpha},b_{o\alpha},c_{o\alpha},\ell_{0,o\alpha}$ for the vectors in both ordered processes. We compare two constraints on these same coefficients. The required output determines the sum of squared SWITCH coefficients and a complex cross term with the residual. Tracing the terminal channel in each causal order will force their corresponding real combination to vanish.
\newpage 
\begin{theorem}\label{thm:one-use}
For any $0<\lambda_A,\lambda_B\le1$, no deterministic QC--CC circuit using each input once implements $\mathcal R_{\lambda_A,\lambda_B}(\mathcal A,\mathcal B)$ for all qubit channels. Equality for $\mathcal A=\Ad_U$ and $\mathcal B=\Ad_V$ for every $U,V\in\mathrm U(2)$ already gives a contradiction.
\end{theorem}
\begin{proof}
Assume that the circuit exists. Its complete output Choi operator $J(U,V)$ from Eq.~\eqref{eq:target-choi} can be evaluated either from the required channel or from the fixed vectors of the circuit. To compare the control sectors while retaining both target legs, define $\Delta_c:P_CP_TF_CF_T\to P_TF_T$ by $\Delta_c\coloneqq\sum_{t,u=0}^1\ket{t,u}\bra{c,t,c,u}$ and define $\Delta\coloneqq\Delta_1-\Delta_0$. Then $\Delta_c\ketket G=\ketket{G_{cc}}$. In particular, $\Delta$ annihilates $\ketket{\id_C\otimes X}$ for every target matrix $X$, whereas $\Delta\ketket{w(U,V)}=\ketket{[U,V]}$ and $\Delta_0\ketket{w(U,V)}=\ketket{VU}$. Writing $J=J(U,V)$, the required channel therefore gives
\begin{align}\label{eq:aggregate-contraction}
 \Delta J\Delta^\dagger=\lambda_A\lambda_B\ketketbra{[U,V]}{[U,V]},\qquad
 \Delta_0J\Delta^\dagger=\lambda_A\lambda_B\ketketbra{VU}{[U,V]}.
\end{align}
In the mixed contraction the three residual terms vanish on the right, even though $\Delta_0$ retains their left factors. For the same contractions, Lemma~\ref{lem:physical} gives $\Delta\ketket{G_{o\alpha}}=e_{o\alpha}\ketket{[U,V]}$ and $\Delta_0\ketket{G_{o\alpha}}=e_{o\alpha}\ketket{VU}+\ketket{H_{o\alpha}}$, with $G_{o\alpha}$ and $H_{o\alpha}$ evaluated at the fixed pair $U,V$. 

\noindent\begin{minipage}{\linewidth}
\noindent Taking the trace compares the required output with the positive decomposition of the same process,
\begin{gather}
 \vcenter{\hbox{\includegraphics{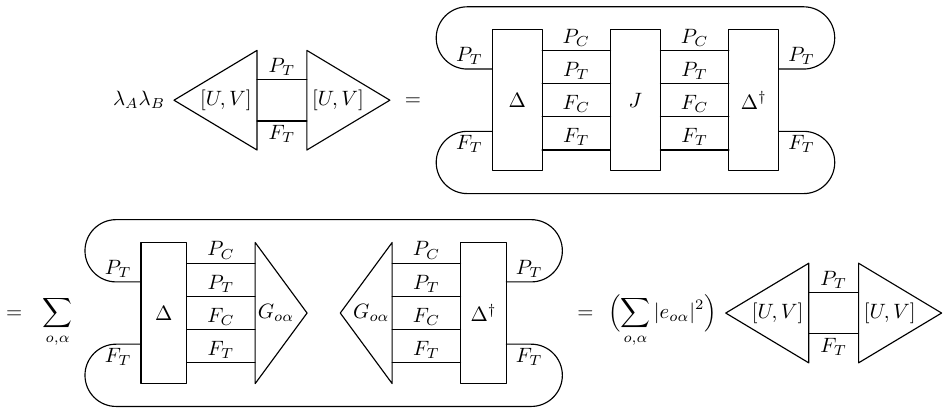}}}\notag\\[5pt]
\begin{aligned}
 \lambda_A\lambda_B\Tr([U,V]^\dagger[U,V])
 &=\Tr(\Delta J\Delta^\dagger)
 =\sum_{o,\alpha}\bigl\|\Delta\ketket{G_{o\alpha}(U,V)}\bigr\|^2=\Tr([U,V]^\dagger[U,V])\sum_{o,\alpha}|e_{o\alpha}|^2.
\end{aligned}\label{eq:weight-calculation}
\end{gather}
\end{minipage}\par
For a noncommuting pair, $\Tr([U,V]^\dagger[U,V])>0$, so
\begin{equation}\label{eq:aggregate-weight}
 \sum_{o,\alpha}|e_{o\alpha}|^2=\lambda_A\lambda_B.
\end{equation}
Substitution into the mixed contraction gives
\begin{align}\label{eq:mixed-contraction}
 \Delta_0J\Delta^\dagger
 &=\sum_{o,\alpha}\bigl(e_{o\alpha}\ketket{VU}+\ketket{H_{o\alpha}(U,V)}\bigr)
                  \overline{e_{o\alpha}}\brabra{[U,V]}\notag\\
 &=\lambda_A\lambda_B\ketketbra{VU}{[U,V]}
   +\left|\sum_{o,\alpha}\overline{e_{o\alpha}}H_{o\alpha}(U,V)\right\rangle\!\rangle\brabra{[U,V]}.
\end{align}
Subtracting the mixed identity in Eq.~\eqref{eq:aggregate-contraction} and applying the difference to $\ketket{[U,V]}$ gives a vector on the two target legs,
\begin{gather*}
 \vcenter{\hbox{\includegraphics{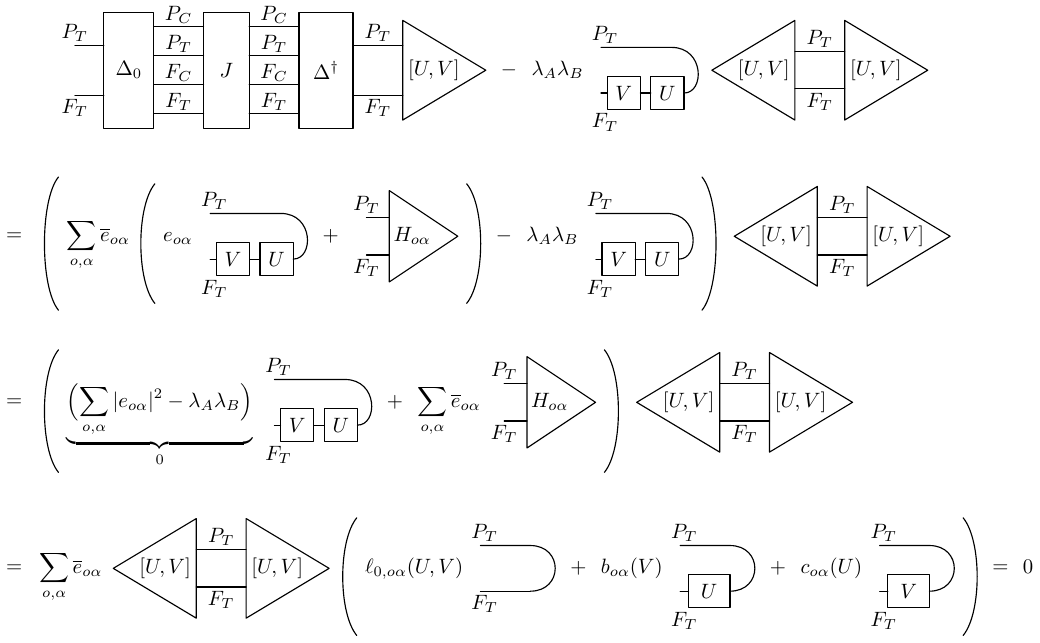}}}\\[5pt]
\begin{aligned}
 &\bigl(\Delta_0J\Delta^\dagger-\lambda_A\lambda_B\ketketbra{VU}{[U,V]}\bigr)\ketket{[U,V]}\\
 &=\Tr([U,V]^\dagger[U,V])\left|\sum_{o,\alpha}\overline e_{o\alpha}H_{o\alpha}(U,V)\right\rangle\!\rangle=0.
\end{aligned}
\end{gather*}
For every noncommuting pair, the strictly positive factor $\Tr([U,V]^\dagger[U,V])$ can be cancelled. Expanding the weighted residual in its fixed normal form gives
\begin{equation}\label{eq:aggregate-residual}
 \left(\sum_{o,\alpha}\overline e_{o\alpha}\ell_{0,o\alpha}(U,V)\right)\id
 +\left(\sum_{o,\alpha}\overline e_{o\alpha}b_{o\alpha}(V)\right)U
 +\left(\sum_{o,\alpha}\overline e_{o\alpha}c_{o\alpha}(U)\right)V=0.
\end{equation}
For noncommuting $U,V$, the matrices $\id,U,V$ are independent. A relation with nonzero $V$ coefficient would express $V$ as an affine function of $U$, and a remaining relation between $\id,U$ would make $U$ scalar. Thus Eq.~\eqref{eq:aggregate-residual} gives $\sum_{o,\alpha}\overline e_{o\alpha}b_{o\alpha}(V)=0$ on these pairs. To reach the identity input, choose $U=\begin{psmallmatrix}0&1\\1&0\end{psmallmatrix}$ and $V_t=\diag(e^{it},e^{-it})$. For real $t$ with $0<|t|<\pi$, the commutator is $[U,V_t]=2i\sin t\begin{psmallmatrix}0&-1\\1&0\end{psmallmatrix}\ne0$. The coefficient therefore vanishes throughout this punctured family. \newpage 
Each $b_{o\alpha}$ is a fixed linear form on all matrices, and there are finitely many of them. Continuity now gives
\begin{equation}\label{eq:aggregate-identity}
 \sum_{o,\alpha}\overline{e_{o\alpha}}b_{o\alpha}(\id)
 =\lim_{t\to0}\sum_{o,\alpha}\overline{e_{o\alpha}}b_{o\alpha}(V_t)=0.
\end{equation}
Equations~\eqref{eq:aggregate-weight} and \eqref{eq:aggregate-identity} fix two combinations of coefficients in the positive decomposition of $W_A+W_B$. To impose causal normalization on these coefficients, we evaluate each $W_o$ from its chronological circuit realization. This representation exposes the terminal channel, whose trace preservation leaves an identity on the last slot output. The resulting constraint on $W_o$ can then be evaluated in the fixed positive decomposition.

For the order $A$ before $B$, include the classical outcome records in the private memory. The initial outcome map is
$\mathcal P_A:\Lop(P)\to\Lop(H_A^IR_A)$. It is CP, whereas the
subsequent maps
$\mathcal D_A:\Lop(H_A^OR_A)\to\Lop(H_B^I\mathcal M_A)$
and $\mathcal N_A:\Lop(H_{B'}^O\mathcal M_A)\to\Lop(F)$ are
channels. Here $H_{B'}^O$ copies the last slot output. Choose Kraus
operators $P_{A,r},D_{A,s},N_{A,\nu}$ for these three maps. As in
Eq.~\eqref{eq:amplitude}, $p,f$ index the input and output controls,
and $t,u$ index the input and output targets. Thus $pt,fu$ are the
complete input and output indices. The indices $i,j,k,l$ still refer to
$H_A^I,H_A^O,H_B^I,H_B^O$,
respectively, while $n,m$ index the private memories $R_A,\mathcal M_A$.
The part of the circuit preceding the terminal channel has amplitude
$\tau^{A}_{rs;ptijkm}\coloneqq\sum_n(D_{A,s})_{km,jn}(P_{A,r})_{in,pt}$ and the full process
amplitude is $C^{A}_{rs\nu;ptijklfu}=\sum_m(N_{A,\nu})_{fu,lm}\tau^{A}_{rs;ptijkm}$.
Inserting arbitrary slot Kraus matrices into this tensor yields the
chronological composition,
\begin{align}\label{eq:physical-prefix-A}
 \sum_{i,j,k,l}C^{A}_{rs\nu;ptijklfu}X_{ji}Y_{lk}
 &=\sum_{i,j,k,l,m,n}(N_{A,\nu})_{fu,lm}Y_{lk}
                  (D_{A,s})_{km,jn}X_{ji}(P_{A,r})_{in,pt}\notag\\
 &=\bigl[N_{A,\nu}(Y\otimes \id_{\mathcal M_A})D_{A,s}
                  (X\otimes \id_{R_A})P_{A,r}\bigr]_{fu,pt}.
\end{align}
Consequently $W_A=\sum_{r,s,\nu}\ket{C^{A}_{rs\nu}}\bra{C^{A}_{rs\nu}}$.
These chronological Kraus vectors and the fixed vectors in Eq.~\eqref{eq:positive-process} are decompositions of the same $W_A$, although their individual terms need not coincide.

Collect these prefix amplitudes into $T_A\coloneqq\sum_{r,s}\ket{\tau^{A}_{rs}}\bra{\tau^{A}_{rs}}$. In terms of physical maps,
$W_A=(\operatorname{id}\otimes\mathcal N_A)
[T_A\otimes\ketketbra \id \id_{H_B^OH_{B'}^O}]$.
$\mathcal N_A$ acts on $H_{B'}^O\mathcal M_A$, and the other
factors retain their labels. The tensor factors are permuted into the
channel's domain before it acts and returned to the order of
Eq.~\eqref{eq:wire-order} afterward.
Trace preservation of the terminal channel now removes the last operation,
\begin{align}\label{eq:terminal-derivation}
 [\Tr_F W_A]_{ptijkl,\,p't'i'j'k'l'}
 &=\sum_{r,s,\nu,f,u}\sum_{m,m'}
       (N_{A,\nu})_{fu,lm}\tau^{A}_{rs;ptijkm}
       \overline{(N_{A,\nu})_{fu,l'm'}\tau^{A}_{rs;p't'i'j'k'm'}}\notag\\
 &=\sum_{m,m'}[T_A]_{ptijkm,\,p't'i'j'k'm'}
       \left\langle l',m'\middle|\sum_\nu N_{A,\nu}^\dagger N_{A,\nu}
                                  \middle|l,m\right\rangle\notag\\
 &=\delta_{ll'}\sum_m[T_A]_{ptijkm,\,p't'i'j'k'm},
\end{align}
where $\sum_\nu N_{A,\nu}^\dagger N_{A,\nu}=\id_{H_{B'}^O\mathcal M_A}$.
The unnormalized identity on $H_B^O$ is therefore independent
of the earlier wires.

For the order $B$ before $A$, use the initial CP map
$\mathcal P_B:\Lop(P)\to\Lop(H_B^IR_B)$ and the channels
$\mathcal D_B:\Lop(H_B^OR_B)\to\Lop(H_A^I\mathcal M_B)$,
$\mathcal N_B:\Lop(H_{A'}^O\mathcal M_B)\to\Lop(F)$.
With Kraus operators $P_{B,r},D_{B,s},N_{B,\nu}$, the amplitudes are
$\tau^{B}_{rs;ptiklm}\coloneqq\sum_n(D_{B,s})_{im,ln}(P_{B,r})_{kn,pt}$ and
$C^{B}_{rs\nu;ptijklfu}=\sum_m(N_{B,\nu})_{fu,jm}\tau^{B}_{rs;ptiklm}$,
where $n,m$ now index $R_B,\mathcal M_B$. The input, output and slot
indices retain their meanings. These amplitudes satisfy
\begin{align}\label{eq:physical-prefix-B}
 \sum_{i,j,k,l}C^{B}_{rs\nu;ptijklfu}X_{ji}Y_{lk}
 &=\sum_{i,j,k,l,m,n}(N_{B,\nu})_{fu,jm}X_{ji}
                        (D_{B,s})_{im,ln}Y_{lk}(P_{B,r})_{kn,pt}\notag\\
 &=\bigl[N_{B,\nu}(X\otimes \id_{\mathcal M_B})D_{B,s}
                         (Y\otimes \id_{R_B})P_{B,r}\bigr]_{fu,pt}.
\end{align}
\newpage 
Define $T_B\coloneqq\sum_{r,s}\ket{\tau^{B}_{rs}}\bra{\tau^{B}_{rs}}$. The corresponding
map representation is
$W_B=(\operatorname{id}\otimes\mathcal N_B)
[T_B\otimes\ketketbra \id \id_{H_A^OH_{A'}^O}]$,
with $\mathcal N_B$ acting on $H_{A'}^O\mathcal M_B$ under the
corresponding permutation of the tensor factors and subsequent return
to the fixed wire order.
Tracing the complete future gives the other terminal condition,
\begin{align}
 [\Tr_F W_B]_{ptijkl,\,p't'i'j'k'l'}
 &=\sum_{r,s,\nu,f,u}\sum_{m,m'}
       (N_{B,\nu})_{fu,jm}\tau^{B}_{rs;ptiklm}
       \overline{(N_{B,\nu})_{fu,j'm'}\tau^{B}_{rs;p't'i'k'l'm'}}\notag\\
 &=\sum_{m,m'}[T_B]_{ptiklm,\,p't'i'k'l'm'}
       \left\langle j',m'\middle|\sum_\nu N_{B,\nu}^\dagger N_{B,\nu}
                                  \middle|j,m\right\rangle\notag\\
 &=\delta_{jj'}\sum_m[T_B]_{ptiklm,\,p't'i'k'l'm}.
\end{align}
Thus, with $Q_o\coloneqq\Tr_{\mathcal M_o}T_o\ge0$, both processes obey the
following identities, where each identity factor is inserted at its
labelled output position in the fixed wire order,
\begin{equation}\label{eq:terminal}
 \Tr_FW_A=\id_{H_B^O}\otimes Q_A,\qquad
 \Tr_FW_B=\id_{H_A^O}\otimes Q_B.
\end{equation}
These conditions apply to the whole ordered process operators. Their initial
effects satisfy $E_A+E_B=\id_P$, so the order probabilities
$\Tr(E_o\rho)$ may depend on the input. In the input-first Choi convention,
$\Tr_{H_o^I R_o}J^{\mathcal P_o}=E_o^T$. Composition with the subsequent
channels preserves these subnormalized traces. This is the two-call specialization of physical QC--CC normalization~\cite{PRXQuantum.2.030335}.

The terminal marginal gives a test of causal normalization. Let $L$ be a Hermitian operator on the factors preceding $F$. If $\Tr_{H_B^O}L=0$, its pairing with $W_A$ vanishes by contracting the last output,
\begin{gather}
 \vcenter{\hbox{\includegraphics{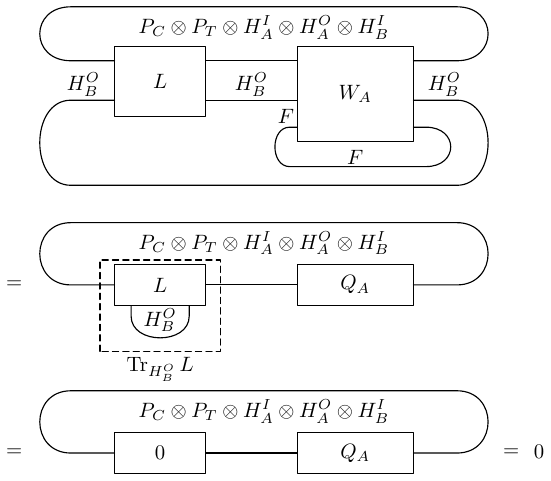}}}\notag\\[5pt]
\begin{aligned}
\Tr[(L\otimes \id_F)W_A]
 &=\Tr[L(\id_{H_B^O}\otimes Q_A)]
 =\Tr[(\Tr_{H_B^O}L)Q_A].
\end{aligned}\label{eq:annihilating-pairing}
\end{gather}
For $W_B$, the corresponding trace is over $H_A^O$. We use these pairings to compare the two control sectors of Eq.~\eqref{eq:physical-vector}. \newpage 
Let $\Pi_a=\ket a\bra a$, $Z=\Pi_0-\Pi_1$ and $\bar a=1-a$. The three residual terms share the control factor $\ketket\id_{P_CF_C}$. Closing both control wires in the outer product of their sum with $Z$ therefore gives
\begin{gather*}
 \vcenter{\hbox{\includegraphics{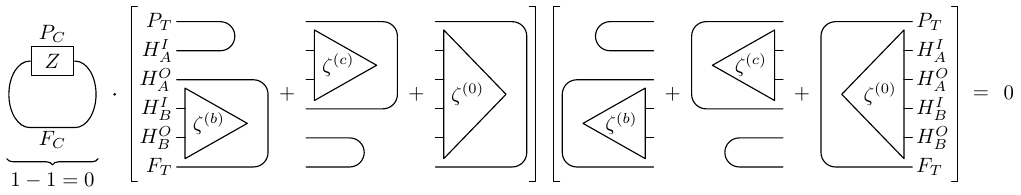}}}\\[5pt]
 \Tr_{P_CF_C}\!\left[(Z_{P_C}\otimes\id)
 (\ketket C-e\ketket S)(\brabra C-\overline e\brabra S)\right]=0.
\end{gather*}
The remaining six spaces are left open, so the cancellation holds before contracting the other factors of either causal test. For the full vector $\ketket C$, the SWITCH contribution and its interference with the residual remain. The target identities select the particular residual value $b(\id)=\zeta^{(b)}_{00}+\zeta^{(b)}_{11}$.

For order $A$ before $B$, restrict the other wires to $t=i=k=a$ and $j=\bar a$. This removes the SWITCH term in control sector $p=0$. At $l=a$, the SWITCH term in sector $p=1$ and the $b$ term both end at $\ket{\bar a}_{F_T}$, while the other residual terms end at $\ket a$. To make this selection traceless on the last output, replace $\Pi_a$ there by $\Pi_a-\id/2=(\Pi_a-\Pi_{\bar a})/2$. The control-sector difference then gives
\begin{equation}\label{eq:causal-test-A}
 L_A=-\frac12\sum_{a=0}^1 Z_{P_C}\otimes\Pi_a^{P_T}\otimes\Pi_a^{H_A^I}
 \otimes\Pi_{\bar a}^{H_A^O}\otimes\Pi_a^{H_B^I}
 \otimes(\Pi_a^{H_B^O}-\Pi_{\bar a}^{H_B^O}).
\end{equation}
The last factor has trace $1-1=0$, so $\Tr[(L_A\otimes\id_F)W_A]=0$ by Eq.~\eqref{eq:annihilating-pairing}. 

\noindent\begin{minipage}{\linewidth}
\noindent To evaluate the same pairing in the fixed positive decomposition, take one vector $\ketket C$ and suppress its order and decomposition indices. Contracting its past and slot legs with the specified basis vectors gives the complete future vector
\begin{gather}
 \vcenter{\hbox{\includegraphics{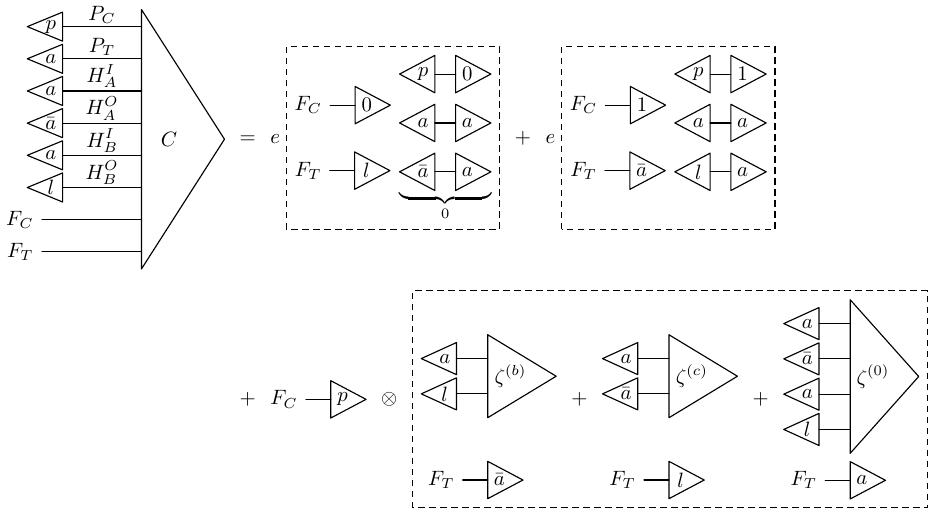}}}\notag\\[5pt]
\begin{aligned}
(\bra{p,a,a,\bar a,a,l}\otimes \id_F)\ketket C
&=\ket p_{F_C}\otimes\bigl[
 \zeta^{(b)}_{al}\ket{\bar a}
 +\zeta^{(c)}_{a\bar a}\ket l
 +\zeta^{(0)}_{a\bar a a l}\ket a
 +e\delta_{p1}\delta_{la}\ket{\bar a}\bigr]_{F_T}.
\end{aligned}\label{eq:A-future-vector}
\end{gather}
\end{minipage}\par
The $p=0$ SWITCH term vanishes because its path requires $j=k$, whereas $j=\bar a$ and $k=a$. Pairing these future vectors with their adjoints and taking the signed trace over the last output gives
\begin{gather}
 \vcenter{\hbox{\includegraphics{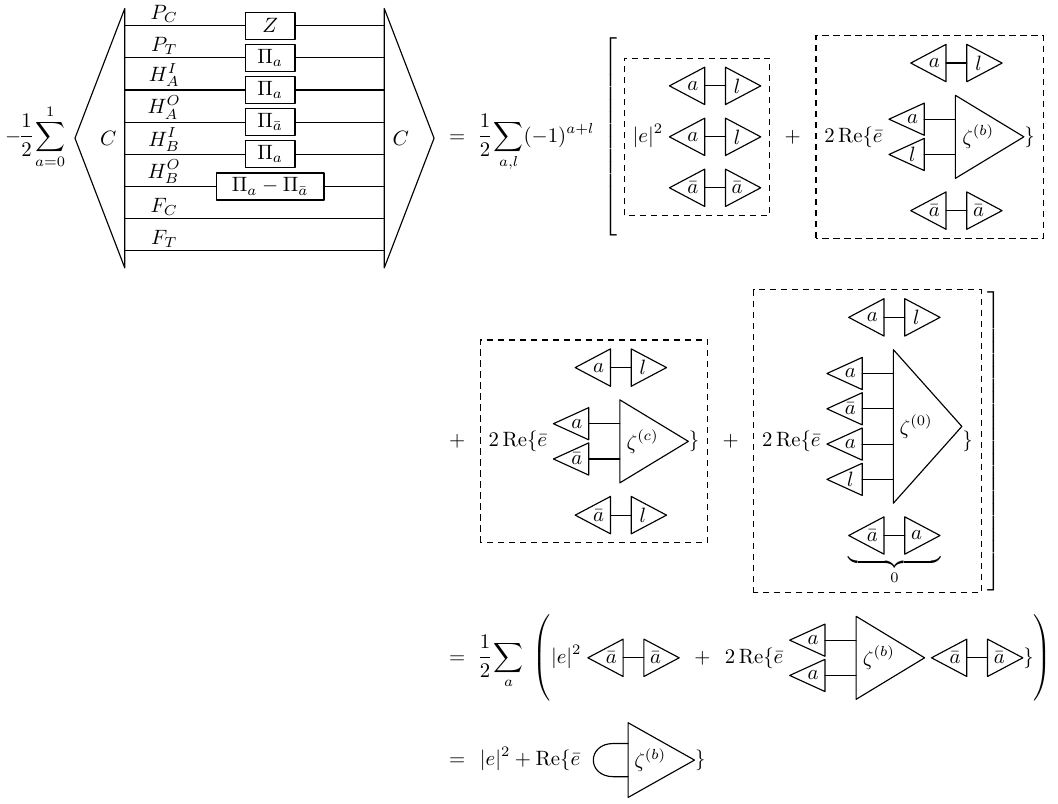}}}\notag\\[5pt]
\begin{aligned}
\brabra C(L_A\otimes \id_F)\ketket C
 &=\frac12\sum_{a,l}(-1)^{a+l}\Bigl[
 \bigl\|\zeta^{(b)}_{al}\ket{\bar a}+\zeta^{(c)}_{a\bar a}\ket l
       +\zeta^{(0)}_{a\bar a a l}\ket a+e\delta_{la}\ket{\bar a}\bigr\|^2-\bigl\|\zeta^{(b)}_{al}\ket{\bar a}+\zeta^{(c)}_{a\bar a}\ket l
       +\zeta^{(0)}_{a\bar a a l}\ket a\bigr\|^2\Bigr]\\
 &=\frac12\sum_{a,l}(-1)^{a+l}\delta_{la}\Bigl[|e|^2
      +2\operatorname{Re}\bigl\{\overline e(\zeta^{(b)}_{al}
        +\zeta^{(c)}_{a\bar a}\delta_{l\bar a}
        +\zeta^{(0)}_{a\bar a a l}\delta_{a\bar a})\bigr\}\Bigr]\\
 &=\frac12\sum_a[|e|^2+2\operatorname{Re}(\overline e\zeta^{(b)}_{aa})]
  =|e|^2+\operatorname{Re}(\overline e b(\id)).
\end{aligned}\label{eq:A-quadratic}
\end{gather}
The squared residual norm is identical in the two control sectors and cancels in their difference. The remaining unwanted overlaps vanish because $\delta_{la}\delta_{l\bar a}=0$ and $\delta_{a\bar a}=0$.

For order $B$ before $A$, the last output is $H_A^O$. To select the same two diagonal coefficients of $b$, take $t=i=1$ and $k=l$, leaving $j$ free. The values $k=0,1$ put the switched contribution in opposite control sectors. Compensating that change by a sign on the two $B$ diagonals gives
\begin{equation}\label{eq:causal-test-B}
 L_B=\frac12Z_{P_C}\otimes\Pi_1^{P_T}\otimes\Pi_1^{H_A^I}
 \otimes Z_{H_A^O}\otimes
 (\Pi_0^{H_B^I}\otimes\Pi_0^{H_B^O}
 -\Pi_1^{H_B^I}\otimes\Pi_1^{H_B^O}).
\end{equation}
Its partial trace over $H_A^O$ vanishes because $\Tr Z=0$. For these wire values the corresponding future vector is
\begin{equation}\label{eq:B-future-vector}
(\bra{p,1,1,j,k,k}\otimes \id_F)\ketket C
=\ket p_{F_C}\otimes\bigl[
 \zeta^{(b)}_{kk}\ket j
 +\zeta^{(c)}_{1j}\delta_{k1}\ket k
 +\zeta^{(0)}_{1jkk}\ket1
 +e\delta_{p0}\delta_{jk}\ket k
 +e\delta_{p1}\delta_{k1}\ket j\bigr]_{F_T}.
\end{equation}
Here the two SWITCH paths survive for different values of the free indices. Taking the signed difference of the squared norms of the complete future vectors gives
\begin{gather}
 \vcenter{\hbox{\includegraphics{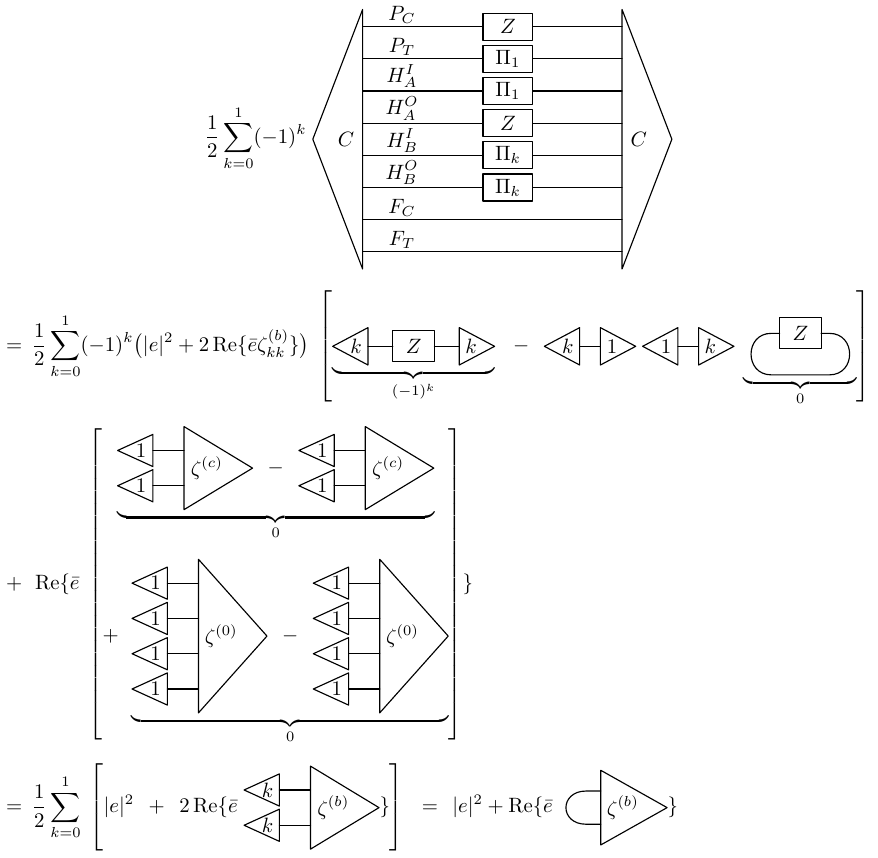}}}\notag\\[5pt]
\begin{aligned}
\brabra C(L_B\otimes \id_F)\ketket C
 &=\frac12\sum_{j,k}(-1)^{j+k}\Bigl[
 \bigl\|\zeta^{(b)}_{kk}\ket j+\zeta^{(c)}_{1j}\delta_{k1}\ket k
       +\zeta^{(0)}_{1jkk}\ket1+e\delta_{jk}\ket k\bigr\|^2\\
 &\hspace{2em}-\bigl\|\zeta^{(b)}_{kk}\ket j+\zeta^{(c)}_{1j}\delta_{k1}\ket k
       +\zeta^{(0)}_{1jkk}\ket1+e\delta_{k1}\ket j\bigr\|^2\Bigr]\\
 &=\frac12\sum_{j,k}(-1)^{j+k}\Bigl[|e|^2(\delta_{jk}-\delta_{k1})
       +2\operatorname{Re}\Bigl\{\overline e\Bigl(
            \zeta^{(b)}_{kk}(\delta_{jk}-\delta_{k1})\\
 &\hspace{2em}+\zeta^{(c)}_{1j}(\delta_{jk}\delta_{k1}-\delta_{k1}\delta_{jk})
       +\zeta^{(0)}_{1jkk}(\delta_{jk}\delta_{k1}-\delta_{k1}\delta_{j1})
       \Bigr)\Bigr\}\Bigr]\\
 &=\frac12\sum_{j,k}(-1)^{j+k}\delta_{j0}(-1)^k
       [|e|^2+2\operatorname{Re}(\overline e\zeta^{(b)}_{kk})]\\
 &=|e|^2+\operatorname{Re}(\overline e b(\id)).
\end{aligned}\label{eq:B-quadratic}
\end{gather}
Again the squared residual norms cancel. The cross term involving $c$ vanishes identically. Both products of Kronecker deltas multiplying $\zeta^{(0)}$ require $j=k=1$, so their difference vanishes as well. Finally $\delta_{jk}-\delta_{k1}=\delta_{j0}(-1)^k$ leaves precisely the two diagonal $b$ coefficients.

The terminal trace calculation establishes $\Tr[(L_o\otimes\id_F)W_o]=0$ for each order. Evaluating the same pairing in the fixed positive decomposition and using Eqs.~\eqref{eq:A-quadratic} and \eqref{eq:B-quadratic} therefore gives
\begin{align}\label{eq:causal-sums}
 0=\Tr[(L_A\otimes \id_F)W_A]
   =\sum_\alpha[|e_{A\alpha}|^2+\operatorname{Re}(\overline e_{A\alpha}b_{A\alpha}(\id))],\quad
 0=\Tr[(L_B\otimes \id_F)W_B]
   =\sum_\alpha[|e_{B\alpha}|^2+\operatorname{Re}(\overline e_{B\alpha}b_{B\alpha}(\id))].
\end{align}
Adding these equations and using Eqs.~\eqref{eq:aggregate-weight} and \eqref{eq:aggregate-identity} yields
\begin{equation}\label{eq:contradiction}
 0=\sum_{o,\alpha}|e_{o\alpha}|^2
   +\operatorname{Re}\sum_{o,\alpha}\overline e_{o\alpha}b_{o\alpha}(\id)
   =\lambda_A\lambda_B>0.
\end{equation}
\end{proof}

For finite $M,N\geq1$, Corollary~\ref{cor:positive-average} gives $\lambda_A=1/M$ and $\lambda_B=1/N$. Their product is positive, so the contradiction excludes exact universal simulation for every such qubit query allocation.

\subsection{Dimension transfer and omitted inputs}

\begin{proposition}\label{prop:supp_dimension_transfer}
For $d>2$, an exact universal QC--CC simulator of $\mathsf S_d^{(2)}$
with positive finite counts $M,N$ would give a qubit simulator with the same counts.
\end{proposition}
\begin{proof}
An arbitrary simulator can send a query outside any chosen qubit code, even when its external target input is encoded in that code. We therefore need to implement a channel on the entire $d$-dimensional query space using one genuine qubit call. Choose an isometry $V_{\rm code}:\C^2\to\C^d$, with projections $\Pi_{\rm code}=V_{\rm code}V_{\rm code}^\dagger$ and $\Pi_\perp=\id_d-\Pi_{\rm code}$. Perform the projective measurement onto the code and its complement, recording the outcome in private memory. On the code outcome, decode the query, apply the unknown qubit channel once, and re-encode its output. On the complement outcome, store the projected query, apply the unknown channel to a fixed dummy qubit, discard that output, and return the stored system. Both outcomes use exactly one oracle call.

Let $Q\simeq\C^d$ be the incoming query and $Q'\simeq\C^2$ the oracle input. A private register $R_{\rm st}\simeq\C^d$ stores the component in the complementary subspace, while $R$ includes all other registers and any reference. With a fixed qubit state $\tau$ and unit vector $\ket0_{R_{\rm st}}$, the initial instrument has output factors $Q'R_{\rm st}R$. Products below are ordered by their subsystem labels, so the first line includes the fixed permutation from $Q'R R_{\rm st}$ to $Q'R_{\rm st}R$. The instrument acts as
\begin{align}\label{eq:wrapper-instrument}
 \mathcal N_{\rm code}(X_{QR})
 &=\bigl[(V_{\rm code}^\dagger\otimes \id_R)X(V_{\rm code}\otimes \id_R)\bigr]_{Q'R}
       \otimes\ket0\bra0_{R_{\rm st}},\notag\\
 \mathcal N_\perp(X_{QR})
 &=\tau_{Q'}\otimes\bigl[(\Pi_\perp\otimes \id_R)X(\Pi_\perp\otimes \id_R)\bigr]_{R_{\rm st}R},\notag\\
 \Tr_{Q'R_{\rm st}}(\mathcal N_{\rm code}+\mathcal N_\perp)(X)
 &=\Tr_Q[X((\Pi_{\rm code}+\Pi_\perp)\otimes \id_R)]=\Tr_Q X.
\end{align}
The outcome is recorded in the private memory. Applying $\mathcal A$ to $Q'$ and returning the output according to the recorded outcome gives a channel on the entire query space
\begin{align}\label{eq:wrapper-reference}
 (\widetilde{\mathcal A}\otimes\operatorname{id}_{\Lop(R)})(X)
 &=(V_{\rm code}\otimes \id_R)(\mathcal A\otimes\operatorname{id}_{\Lop(R)})
   [(V_{\rm code}^\dagger\otimes \id_R)X(V_{\rm code}\otimes \id_R)]
      (V_{\rm code}^\dagger\otimes \id_R)\notag\\
 &\quad+\Tr_{Q'}[\mathcal A(\tau)_{Q'}\otimes
                 (\Pi_\perp\otimes \id_R)X(\Pi_\perp\otimes \id_R)]\notag\\
 &=(V_{\rm code}\otimes \id_R)(\mathcal A\otimes\operatorname{id}_{\Lop(R)})
   [(V_{\rm code}^\dagger\otimes \id_R)X(V_{\rm code}\otimes \id_R)]
      (V_{\rm code}^\dagger\otimes \id_R)\notag\\
 &\quad+(\Pi_\perp\otimes \id_R)X(\Pi_\perp\otimes \id_R).
\end{align}
Thus $\widetilde{\mathcal A}(X)=V_{\rm code}\mathcal A(V_{\rm code}^\dagger XV_{\rm code})V_{\rm code}^\dagger+\Pi_\perp X\Pi_\perp$. For a Kraus family $\{A_\alpha\}_{\alpha\in\mathsf A}$ of the qubit channel, it has Kraus operators $A'_\alpha\coloneqq V_{\rm code}A_\alpha V_{\rm code}^\dagger$ and $A'_\perp\coloneqq\Pi_\perp$, which obey
$
  \sum_\alpha A_\alpha'^\dagger A'_\alpha+A_\perp'^\dagger A'_\perp
 =V_{\rm code}\id_2V_{\rm code}^\dagger+\Pi_\perp=\id_d.
$
For $\{B_\beta\}_{\beta\in\mathsf B}$, the same construction defines $\widetilde{\mathcal B}$. In particular, the embedded qubit identity is the pinching onto the code and its orthogonal complement $\widetilde{\mathcal{I}}(X)=\Pi_{\rm code}X\Pi_{\rm code}+\Pi_\perp X\Pi_\perp$, rather than the identity channel on the entire $d$-dimensional space.

We next restrict the ideal complete SWITCH to the qubit code. Its external target is encoded by $\mathcal V(X)=V_{\rm code}XV_{\rm code}^\dagger$ and decoded by
\begin{equation}\label{eq:decoder}
 \mathcal D(Y)=V_{\rm code}^\dagger YV_{\rm code}+\Tr(\Pi_\perp Y)\tau.
\end{equation}
For $\tau=\sum_a\tau_a\ket a\bra a$ and an orthonormal complement basis $\{\ket m\}$, this decoder has Kraus operators $D_\star\coloneqq V_{\rm code}^\dagger$ and $D_{am}\coloneqq\sqrt{\tau_a}\ket a\bra m$. They satisfy
\begin{align}\label{eq:decoder-completeness}
 D_\star^\dagger D_\star+\sum_{a,m}D_{am}^\dagger D_{am}
 =\Pi_{\rm code}+\sum_m\ket m\bra m=\id_d,\qquad
 D_\star V_{\rm code}=\id_2,\qquad D_{am}V_{\rm code}=0.
\end{align}
These formulas give a CPTP decoder on the entire output space. The target encoder $\operatorname{id}_{\Lop(P_C)}\otimes\mathcal V$ and decoder $\operatorname{id}_{\Lop(F_C)}\otimes\mathcal D$ change only the target dimension, leaving the input and output control registers intact.

To calculate the complete encoded target, recall $K_{0,\alpha\beta}=B_\beta A_\alpha$ and $K_{1,\alpha\beta}=A_\alpha B_\beta$ for $\alpha\in\mathsf A$, $\beta\in\mathsf B$. Adjoin a new complement label to each index set, giving $\widehat{\mathsf A}=\mathsf A\sqcup\{\perp\}$ and $\widehat{\mathsf B}=\mathsf B\sqcup\{\perp\}$. For $\widehat\alpha\in\widehat{\mathsf A}$ and $\widehat\beta\in\widehat{\mathsf B}$, define the embedded products $K'_{0,\widehat\alpha\widehat\beta}=B'_{\widehat\beta}A'_{\widehat\alpha}$ and $K'_{1,\widehat\alpha\widehat\beta}=A'_{\widehat\alpha}B'_{\widehat\beta}$. Their products with the encoder satisfy
\begin{equation}\label{eq:encoded-products}
 K'_{c,\widehat\alpha\widehat\beta}V_{\rm code}
 =\begin{cases}
 V_{\rm code}K_{c,\alpha\beta},&\widehat\alpha=\alpha\in\mathsf A,\ \widehat\beta=\beta\in\mathsf B,\\
 0,&\widehat\alpha=\perp\text{ or }\widehat\beta=\perp.
 \end{cases}
\end{equation}
Indeed a code Kraus product remains in the code, while $\Pi_\perp V_{\rm code}=0$ and $\Pi_\perp A'_\alpha=A'_\alpha\Pi_\perp=0$, with the same identities for $B'_\beta$. For any $c,c'\in\{0,1\}$ and arbitrary $X_{TR}$, expand all oracle and decoder Kraus indices. Letting $\gamma$ range over all decoder Kraus operators gives
\begin{align}\label{eq:encoded-blocks}
 &[(\operatorname{id}_{\Lop(F_C)}\otimes\mathcal D\otimes\operatorname{id}_{\Lop(R)})\circ
   (\SW_d^{(2)}(\widetilde{\mathcal A},\widetilde{\mathcal B})\otimes\operatorname{id}_{\Lop(R)})
   \circ(\operatorname{id}_{\Lop(P_C)}\otimes\mathcal V\otimes\operatorname{id}_{\Lop(R)})]
       (\ket c\bra{c'}\otimes X)\notag\\
 &=\ket c\bra{c'}\otimes\sum_{\gamma,\widehat\alpha,\widehat\beta}
 (D_\gamma K'_{c,\widehat\alpha\widehat\beta}V_{\rm code}\otimes \id_R)X
 (D_\gamma K'_{c',\widehat\alpha\widehat\beta}V_{\rm code}\otimes \id_R)^\dagger\notag\\
 &=\ket c\bra{c'}\otimes\sum_{\alpha,\beta}
 (D_\star V_{\rm code}K_{c,\alpha\beta}\otimes \id_R)X
 (D_\star V_{\rm code}K_{c',\alpha\beta}\otimes \id_R)^\dagger\notag\\
 &=\ket c\bra{c'}\otimes\sum_{\alpha,\beta}
 (K_{c,\alpha\beta}\otimes \id_R)X(K_{c',\alpha\beta}^\dagger\otimes \id_R).
\end{align}
Equation~\eqref{eq:encoded-products} removes every term containing an oracle complement index, and Eq.~\eqref{eq:decoder-completeness} removes every remaining decoder complement term. The surviving terms have the same pair $\alpha,\beta$ on both sides of $X$, so the identity includes control interference as well as the evolved target.

The assumed simulator is exact for every pair of $d$-dimensional channels, hence for $\widetilde{\mathcal A},\widetilde{\mathcal B}$. Precomposing it with the target encoder and postcomposing it with the target decoder therefore gives
\begin{align}\label{eq:encoded-switch}
 (\operatorname{id}_{\Lop(F_C)}\otimes\mathcal D)\circ
 \mathcal C_{M,N}[\widetilde{\mathcal A},\widetilde{\mathcal B}]
 \circ(\operatorname{id}_{\Lop(P_C)}\otimes\mathcal V)
 =(\operatorname{id}_{\Lop(F_C)}\otimes\mathcal D)\circ
 \SW_d^{(2)}(\widetilde{\mathcal A},\widetilde{\mathcal B})
 \circ(\operatorname{id}_{\Lop(P_C)}\otimes\mathcal V)=\SW_2^{(2)}(\mathcal A,\mathcal B).
\end{align}
The left-hand side is implemented by replacing each original call with the instrument in Eq.~\eqref{eq:wrapper-instrument}, one qubit query, and the return operation. Its classical outcomes become part of the existing records, and $R_{\rm st}$ becomes part of the private quantum memory. Every original occurrence still calls the same input once on every branch. The resulting qubit circuit is therefore QC--CC with exactly $M,N$ queries. Applying Corollary~\ref{cor:positive-average} to that circuit and its qubit identity mixtures contradicts Theorem~\ref{thm:one-use}.
\end{proof}

\begin{lemma}\label{lem:supp_omitted_binary}
For every $d\geq2$, an exact universal two-channel SWITCH simulator must call both inputs.
\end{lemma}
\begin{proof}
Let $U_{01}$ exchange $\ket0,\ket1$ and fix their orthogonal complement,
and set $\rho=\ket0\bra0_{P_C}\otimes\ket0\bra0_{P_T}$. With $\mathcal{I}$
the identity channel and $\mathcal U_{01}(X)=U_{01}XU_{01}^\dagger$, the two omitted-input tests are
\begin{align}\label{eq:supp_omitted_binary_outputs}
\mathsf S_d^{(2)}(\mathcal{I},\mathcal{I})(\rho)=\ket{00}\bra{00}_{F_CF_T},\qquad
\mathsf S_d^{(2)}(\mathcal U_{01},\mathcal{I})(\rho)
=\mathsf S_d^{(2)}(\mathcal{I},\mathcal U_{01})(\rho)=\ket{01}\bra{01}_{F_CF_T}.
\end{align}
If $M=0$, the circuit has identical outputs on the first and second input pairs.
If $N=0$, it has identical outputs on the first and third pairs.
Each equality conflicts with the orthogonal target outputs above. This includes $M=N=0$.
\end{proof}

\begin{theorem}[Exact universal two-channel obstruction]\label{thm:supp_all_finite_binary}
For every integer $d\geq2$ and every fixed finite pair $M,N\in\mathbb Z_{\geq0}$,
no deterministic QC--CC circuit satisfies Eq.~\eqref{eq:supp_exact_two_label_task}
for all $\mathcal A,\mathcal B\in\CPTP_d$. The equality concerns the complete
channel from $P_C\otimes P_T$ to $F_C\otimes F_T$, including every reference extension.
\end{theorem}
\begin{proof}
For $d=2$ and $M,N>0$, the physical average in
Corollary~\ref{cor:positive-average} contradicts Theorem~\ref{thm:one-use}.
Proposition~\ref{prop:supp_dimension_transfer} reduces every positive-count
case with $d>2$ to that qubit contradiction without changing the counts.
Lemma~\ref{lem:supp_omitted_binary} excludes the remaining zero-count cases.
\end{proof}

\section{The complete multichannel switch}
\label{sec:supp_complete_r}
\label{sec:supp_complete_switch}

The complete $r$-switch coherently controls all $r!$ orders of $r$ labelled channels. We first describe its physical wiring and the complete channel it induces. Two permutation sectors differing by an adjacent exchange then yield the two-channel SWITCH when the remaining inputs are identities. Universality transfers this channel identity to a hypothetical simulator. The retention construction in S2 then absorbs its identity calls while preserving every occurrence of the selected pair, giving the two-channel contradiction for any fixed finite multichannel inventory.

	\subsection{Complete \texorpdfstring{$r$}{r}-switch channel and permutation coherence}
	\label{subsec:supp_complete_switch_channel}
	
	Fix integers $r,d\geq2$, write $[r]=\{1,\ldots,r\}$, and let
	$\mathfrak S_r$ be the symmetric group. The control input and output spaces
	are independent copies $P_C^{(r)}\simeq F_C^{(r)}\simeq\C^{r!}$, with
	$\{\lvert\pi\rangle:\pi\in\mathfrak S_r\}$ orthonormal,
	whereas $P_T^{(r)}\simeq F_T^{(r)}\simeq\C^d$. Let
	$P^{(r)}=P_C^{(r)}\otimes P_T^{(r)}$ and
	$F^{(r)}=F_C^{(r)}\otimes F_T^{(r)}$.  For each label
	$\ell\in[r]$, choose an arbitrary finite Kraus family
	\begin{equation}
	\mathcal A_\ell(\rho)
	=\sum_{a_\ell=1}^{n_\ell}
	A^{(\ell)}_{a_\ell}\rho A^{(\ell)\dagger}_{a_\ell},
	\;
	\sum_{a_\ell=1}^{n_\ell}
	A^{(\ell)\dagger}_{a_\ell}A^{(\ell)}_{a_\ell}=\id_d.
	\end{equation}
	Our temporal convention is that the tuple
	$\pi=(\pi(1),\ldots,\pi(r))$ lists the calls from first to last.  For the
	multi-index $\boldsymbol a=(a_1,\ldots,a_r)$, define
	\begin{align}
		V_{\pi,\boldsymbol a}
		\coloneqq
		A^{(\pi(r))}_{a_{\pi(r)}}\cdots
		A^{(\pi(2))}_{a_{\pi(2)}}
		A^{(\pi(1))}_{a_{\pi(1)}},
		\qquad
		W_{\boldsymbol a}
		\coloneqq
		\sum_{\pi\in\mathfrak S_r}
		\lvert\pi\rangle_{F_C^{(r)}}
		\!\langle\pi\rvert_{P_C^{(r)}}\otimes
		V_{\pi,\boldsymbol a}.
		\label{eq:supp_complete_switch_kraus}
	\end{align}

For any ordering of $\mathfrak S_r$, $W_{\boldsymbol a}$ is block diagonal on the corresponding direct sums of target spaces, with blocks $V_{\pi,\boldsymbol a}$. Its products act from right to left, so the first temporal call appears on the right. The same Kraus multi-index $\boldsymbol a$ occurs in every control sector.

Write $I_\ell\simeq O_\ell\simeq\C^d$ for the input and output wires of the slot carrying label $\ell$.

	Although each controlled Kraus operator is block diagonal in the permutation
	basis, a coherent control input can produce off-diagonal output blocks.

The process-vector representation of the quantum $N$-switch in Ref.~\cite{PRXQuantum.2.030335} fixes the wiring of each permutation sector. Using S1.A's input-first vectorization, the complete $r$-switch process vector can be written with its labelled factors grouped by wire connections. The fixed tensor permutation placing those factors in S1.A's order is understood,

	\begin{equation}
		\ketket{w_{\rm sw}^{(r)}}
		=\sum_{\pi\in\mathfrak S_r}
		\lvert\pi\rangle_{P_C^{(r)}}
		\otimes\lvert\pi\rangle_{F_C^{(r)}}
		\otimes\ketket{\id}^{P_T^{(r)}I_{\pi(1)}}
		\otimes
		\bigotimes_{t=1}^{r-1}
		\ketket{\id}^{O_{\pi(t)}I_{\pi(t+1)}}
		\otimes\ketket{\id}^{O_{\pi(r)}F_T^{(r)}}.
		\label{eq:supp_complete_switch_process_vector}
	\end{equation}
	For a fixed $\pi$ and Kraus multi-index $\boldsymbol a$, contraction
	with the labelled slot operators follows the temporal order in
	Eq.~\eqref{eq:supp_complete_switch_kraus}, namely
	\begin{align}
		&\ketket{\id}^{P_T^{(r)}I_{\pi(1)}}*
		\ketket{A^{(\pi(1))}_{a_{\pi(1)}}}^{I_{\pi(1)}O_{\pi(1)}}*
		\ketket{\id}^{O_{\pi(1)}I_{\pi(2)}}*
		\ketket{A^{(\pi(2))}_{a_{\pi(2)}}}^{I_{\pi(2)}O_{\pi(2)}}*
		\cdots *
		\ketket{\id}^{O_{\pi(r-1)}I_{\pi(r)}}*
		\ketket{A^{(\pi(r))}_{a_{\pi(r)}}}^{I_{\pi(r)}O_{\pi(r)}}*
		\ketket{\id}^{O_{\pi(r)}F_T^{(r)}}
		\nonumber\\
		&\hspace{5em}=
		\ketket{A^{(\pi(r))}_{a_{\pi(r)}}\cdots
			A^{(\pi(1))}_{a_{\pi(1)}}}^{P_T^{(r)}F_T^{(r)}}
		=\ketket{V_{\pi,\boldsymbol a}}^{P_T^{(r)}F_T^{(r)}}.
		\label{eq:supp_complete_switch_general_process_link}
	\end{align}

For $r=3$ and $\pi=231$, the identity wires run from $P_T^{(3)}$ to $I_2$, from $O_2$ to $I_3$, from $O_3$ to $I_1$, and from $O_1$ to $F_T^{(3)}$. The general contraction therefore gives $V_{231,\boldsymbol a}=A^{(1)}_{a_1}A^{(3)}_{a_3}A^{(2)}_{a_2}$, with channel 2 acting first.

	For $r=3$, take the control order
	$(123,132,213,231,312,321)$ and
	$\boldsymbol a=(a_1,a_2,a_3)$. The controlled Kraus operator itself is the
	six-sector block matrix
	\begin{equation}
		\begingroup
		\setlength{\arraycolsep}{1.5pt}
		W_{\boldsymbol a}=
		\begin{pmatrix}
			A^{(3)}_{a_3}A^{(2)}_{a_2}A^{(1)}_{a_1}&0&0&0&0&0\\
			0&A^{(2)}_{a_2}A^{(3)}_{a_3}A^{(1)}_{a_1}&0&0&0&0\\
			0&0&A^{(3)}_{a_3}A^{(1)}_{a_1}A^{(2)}_{a_2}&0&0&0\\
			0&0&0&A^{(1)}_{a_1}A^{(3)}_{a_3}A^{(2)}_{a_2}&0&0\\
			0&0&0&0&A^{(2)}_{a_2}A^{(1)}_{a_1}A^{(3)}_{a_3}&0\\
			0&0&0&0&0&A^{(1)}_{a_1}A^{(2)}_{a_2}A^{(3)}_{a_3}
		\end{pmatrix}.
		\endgroup
		\label{eq:supp_three_switch_products}
	\end{equation}
	The six diagonal products make the temporal convention explicit in the
	control order $(123,132,213,231,312,321)$.
	
	\begin{definition}[Complete $r$-switch]
		\label{def:supp_complete_quantum_switch}
		The complete $r$-switch is the map
		\begin{equation}
			\mathsf S_d^{(r)}(\mathcal A_1,\ldots,\mathcal A_r)(\rho)
			\coloneqq
			\sum_{\boldsymbol a}W_{\boldsymbol a}\rho W_{\boldsymbol a}^{\dagger}.
			\label{eq:supp_complete_switch_channel}
		\end{equation}
	\end{definition}
	Multiplying the two controlled sums in the Kraus expression gives
	\begin{align}
		\mathsf S_d^{(r)}(\mathcal A_1,\ldots,\mathcal A_r)(\rho)
		&=\sum_{\boldsymbol a}
		\left(\sum_{\pi\in\mathfrak S_r}
		\lvert\pi\rangle_{F_C^{(r)}}\!\langle\pi\rvert_{P_C^{(r)}}
		\otimes V_{\pi,\boldsymbol a}\right)\rho
		\left(\sum_{\sigma\in\mathfrak S_r}
		\lvert\sigma\rangle_{P_C^{(r)}}\!\langle\sigma\rvert_{F_C^{(r)}}
		\otimes V_{\sigma,\boldsymbol a}^{\dagger}\right)
		\nonumber\\
		&=\sum_{\pi,\sigma\in\mathfrak S_r}
		\lvert\pi\rangle\!\langle\sigma\rvert_{F_C^{(r)}}
		\otimes
		\sum_{\boldsymbol a}
		V_{\pi,\boldsymbol a}
		\bigl[(\langle\pi\rvert_{P_C^{(r)}}\otimes\id_d)\rho
		(\lvert\sigma\rangle_{P_C^{(r)}}\otimes\id_d)\bigr]
		V_{\sigma,\boldsymbol a}^{\dagger}.
		\label{eq:supp_complete_switch_double_permutation}
	\end{align}
	For example, the $(231,312)$ output-control block is
	\begin{align}
		&(\langle231\rvert_{F_C^{(3)}}\otimes\id_d)
		\mathsf S_d^{(3)}(\mathcal A_1,\mathcal A_2,\mathcal A_3)(\rho)
		(\lvert312\rangle_{F_C^{(3)}}\otimes\id_d)
		\nonumber\\
		&=\sum_{a_1,a_2,a_3}
		A^{(1)}_{a_1}A^{(3)}_{a_3}A^{(2)}_{a_2}
		\bigl[(\langle231\rvert_{P_C^{(3)}}\otimes\id_d)\rho
		(\lvert312\rangle_{P_C^{(3)}}\otimes\id_d)\bigr]
		A^{(3)\dagger}_{a_3}A^{(1)\dagger}_{a_1}A^{(2)\dagger}_{a_2}.
		\label{eq:supp_complete_switch_231_312_block}
	\end{align}

The right-hand product is $V_{312,\boldsymbol a}^{\dagger}$, so its factors occur in the reverse order from $V_{312,\boldsymbol a}$. The shared indices determine this off-diagonal control block. Its value depends on the input channels and can vanish for noisy inputs.

More generally, for any control matrix unit, including a diagonal one, and any operator $X$ on the target and an arbitrary reference $R$, the complete action is
\begin{equation}\label{eq:supp_complete_r_block}
(\mathsf S_d^{(r)}(\mathcal A_1,\ldots,\mathcal A_r)\otimes\operatorname{id}_{\Lop(R)})
(\ket\pi\bra{\pi'}\otimes X)
=\ket\pi\bra{\pi'}\otimes
 \sum_{\boldsymbol a}(V_{\pi,\boldsymbol a}\otimes\id_R)
 X(V_{\pi',\boldsymbol a}^{\dagger}\otimes\id_R).
\end{equation}
The shared Kraus indices determine the cross terms between distinct sectors as well as the channel in each diagonal sector. Tracing out the output control would remove those cross sectors and give a different operation.

To verify that Eq.~\eqref{eq:supp_complete_switch_channel} depends only on the channels, take each $\{A^{(\ell)}_{a_\ell}\}_{a_\ell=1}^{n_\ell}$ to be minimal. Any finite Kraus family $\{B^{(\ell)}_{b_\ell}\}_{b_\ell=1}^{m_\ell}$ of the same channel has the form $B^{(\ell)}_{b_\ell}=\sum_{a_\ell}u^{(\ell)}_{b_\ell a_\ell}A^{(\ell)}_{a_\ell}$ with an $m_\ell\times n_\ell$ isometry.\newline \mbox{Thus $\sum_{b_\ell}\overline{u^{(\ell)}_{b_\ell c_\ell}}u^{(\ell)}_{b_\ell a_\ell}=\delta_{c_\ell a_\ell}$.} Since the coefficients are scalars, multilinearity of every ordered product gives $\widetilde W_{\boldsymbol b}=\sum_{\boldsymbol a}(\prod_{\ell=1}^r u^{(\ell)}_{b_\ell a_\ell})W_{\boldsymbol a}$ with the same coefficients in every control sector. Substituting this transformation into the channel sum contracts the product isometry,

	\begin{equation}
		\sum_{\boldsymbol b}\widetilde W_{\boldsymbol b}\rho
		\widetilde W_{\boldsymbol b}^{\dagger}
		=\sum_{\boldsymbol a,\boldsymbol c}
		\left[
		\prod_{\ell=1}^{r}
		\sum_{b_\ell}u^{(\ell)}_{b_\ell a_\ell}
		\overline{u^{(\ell)}_{b_\ell c_\ell}}
		\right]
		W_{\boldsymbol a}\rho W_{\boldsymbol c}^{\dagger}
		=\sum_{\boldsymbol a,\boldsymbol c}
		\left(\prod_{\ell=1}^{r}\delta_{a_\ell c_\ell}\right)
		W_{\boldsymbol a}\rho W_{\boldsymbol c}^{\dagger}
		=\sum_{\boldsymbol a}W_{\boldsymbol a}\rho W_{\boldsymbol a}^{\dagger}.
		\label{eq:supp_complete_switch_kraus_independence}
	\end{equation}

Every finite representation is related to the same minimal family, so this identity also compares any two finite representations. Tensoring the calculation with an identity on a reference preserves the equality.

The Kraus form gives complete positivity. For a fixed temporal order $\pi$, sum the last temporal Kraus index first in $V_{\pi,\boldsymbol a}^{\dagger}V_{\pi,\boldsymbol a}$. The corresponding completeness relation removes its two adjacent factors. Repeating this backwards through the order gives
\begin{align}\label{eq:supp_complete_r_nested_tp}
 \sum_{\boldsymbol a}V_{\pi,\boldsymbol a}^{\dagger}V_{\pi,\boldsymbol a}
 &=\sum_{a_{\pi(1)},\ldots,a_{\pi(r-1)}}
 A^{(\pi(1))\dagger}_{a_{\pi(1)}}\cdots A^{(\pi(r-1))\dagger}_{a_{\pi(r-1)}}
 \left(\sum_{a_{\pi(r)}}A^{(\pi(r))\dagger}_{a_{\pi(r)}}A^{(\pi(r))}_{a_{\pi(r)}}\right)
 A^{(\pi(r-1))}_{a_{\pi(r-1)}}\cdots A^{(\pi(1))}_{a_{\pi(1)}}\notag\\
 &=\sum_{a_{\pi(1)},\ldots,a_{\pi(r-1)}}
 A^{(\pi(1))\dagger}_{a_{\pi(1)}}\cdots A^{(\pi(r-1))\dagger}_{a_{\pi(r-1)}}
 A^{(\pi(r-1))}_{a_{\pi(r-1)}}\cdots A^{(\pi(1))}_{a_{\pi(1)}}\notag\\
 &=\cdots=\sum_{a_{\pi(1)}}A^{(\pi(1))\dagger}_{a_{\pi(1)}}A^{(\pi(1))}_{a_{\pi(1)}}=\id_d.
\end{align}

	For $r=3$ and $\pi=231$, the operator order is visible without
	recurrence notation as
	\begin{align}
		\sum_{a_1,a_2,a_3}V_{231,\boldsymbol a}^{\dagger}
		V_{231,\boldsymbol a}
		&=\sum_{a_1,a_2,a_3}
		A^{(2)\dagger}_{a_2}A^{(3)\dagger}_{a_3}A^{(1)\dagger}_{a_1}
		A^{(1)}_{a_1}A^{(3)}_{a_3}A^{(2)}_{a_2}
		=\sum_{a_2,a_3}A^{(2)\dagger}_{a_2}A^{(3)\dagger}_{a_3}
		\left(\sum_{a_1}A^{(1)\dagger}_{a_1}A^{(1)}_{a_1}\right)
		A^{(3)}_{a_3}A^{(2)}_{a_2}
		\nonumber\\
		&=\sum_{a_2}A^{(2)\dagger}_{a_2}
		\left(\sum_{a_3}A^{(3)\dagger}_{a_3}A^{(3)}_{a_3}\right)
		A^{(2)}_{a_2}
		=\sum_{a_2}A^{(2)\dagger}_{a_2}A^{(2)}_{a_2}=\id_d.
		\label{eq:supp_complete_switch_231_tp}
	\end{align}
	Orthogonality of the control projectors now yields
	\begin{equation}
		\sum_{\boldsymbol a}W_{\boldsymbol a}^{\dagger}W_{\boldsymbol a}
		=\sum_{\pi\in\mathfrak S_r}
		\lvert\pi\rangle\!\langle\pi\rvert_{P_C^{(r)}}\otimes
		\sum_{\boldsymbol a}V_{\pi,\boldsymbol a}^{\dagger}V_{\pi,\boldsymbol a}
		=\sum_{\pi\in\mathfrak S_r}
		\lvert\pi\rangle\!\langle\pi\rvert_{P_C^{(r)}}\otimes\id_d
		=\id_{P^{(r)}}.
		\label{eq:supp_complete_switch_control_tp}
	\end{equation}
	Thus Definition~\ref{def:supp_complete_quantum_switch} is CPTP for arbitrary
	finite Kraus families.
	
	Equation~\eqref{eq:supp_complete_switch_general_process_link} shows directly
	that filling every labelled slot produces the ordered products in
	Eq.~\eqref{eq:supp_complete_switch_channel}.  This QC--QC realization uses
	one call to each labelled channel, for total query count $r$.  Its
	$r!$-dimensional control and coherent routing are physical resources
	separate from the black-box call count.  Every adjacent pair of permutation
	sectors supplies the two coherent orders used in the reduction below.

\subsection{Extracting two channels from the operation and its simulator}

Fix distinct labels $i,j$. Choose permutations $\pi_0,\pi_1$ differing by
an adjacent exchange of these labels, with $i$ before $j$ in $\pi_0$.
Fill the other channels with their identity channels. Only the two
Kraus indices $a,b$ remain, and the selected products are
$V_{\pi_0,ab}=A^{(j)}_bA^{(i)}_a$ and $V_{\pi_1,ab}=A^{(i)}_aA^{(j)}_b$.

Let $J_{ij}^{\rm in}$ and $J_{ij}^{\rm out}$ encode a qubit into the input
and output control bases by $\ket c\mapsto\ket{\pi_c}$. The channel
$\mathcal E_{ij}$ has the single Kraus operator $J_{ij}^{\rm in}\otimes \id_d$.
Define a decoder on the entire output space by the Kraus operators
\begin{align}\label{eq:supp_face_decoder}
D_\star&=J_{ij}^{{\rm out}\dagger}\otimes \id_d,
\qquad D_\mu=\ket0\bra\mu\otimes \id_d
\quad(\mu\notin\{\pi_0,\pi_1\}),\\
D_\star^\dagger D_\star+\sum_\mu D_\mu^\dagger D_\mu
&=\bigl[J_{ij}^{\rm out}J_{ij}^{{\rm out}\dagger}
 +(\id-J_{ij}^{\rm out}J_{ij}^{{\rm out}\dagger})\bigr]\otimes \id_d=\id.
\end{align}
Thus $\mathcal D_{ij}$ is CPTP and acts as the identity on the target.
On the selected control subspace it deterministically decodes the qubit.
For $r=2$ the complement is empty.

\begin{proposition}\label{prop:supp_complete_pair_reduction}
The composition of the encoder $\mathcal E_{ij}$, the complete $r$-switch with all unselected inputs set to identity, and the decoder $\mathcal D_{ij}$ is the complete two-channel SWITCH.
If a deterministic universal QC--CC simulator of $\mathsf S_d^{(r)}$ has finite counts $\boldsymbol K=(K_1,\ldots,K_r)$, perform this encoding and decoding on the simulator and fill every occurrence of the other labels with identity channels. For $K_i+K_j>0$, absorbing those identity calls by Proposition~\ref{prop:extraction} gives a deterministic QC--CC two-channel simulator with exactly $K_i,K_j$ calls. Both channel identities hold on arbitrary reference extensions.
\end{proposition}
\begin{proof}
For the ideal operation, the encoding selects the two permutation sectors, and the identity fillings contract to straight target wires. The same Kraus indices $a,b$ occur in both terms, giving
\begin{equation*}
 \vcenter{\hbox{\includegraphics{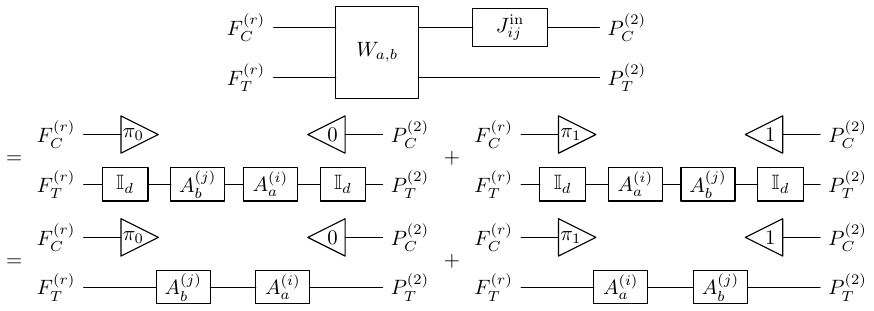}}}
\end{equation*}
\begin{align}\label{eq:supp_face_kraus}
W_{a,b}(J_{ij}^{\rm in}\otimes \id_d)
 &=\ket{\pi_0}\bra0\otimes V_{\pi_0,ab}
   +\ket{\pi_1}\bra1\otimes V_{\pi_1,ab},\\
D_\star W_{a,b}(J_{ij}^{\rm in}\otimes \id_d)
 &=\ket0\bra0\otimes V_{\pi_0,ab}+\ket1\bra1\otimes V_{\pi_1,ab},\\
D_\mu W_{a,b}(J_{ij}^{\rm in}\otimes \id_d)&=0.
\end{align}
For any $c,c'\in\{0,1\}$ and target and reference operator $X$, expand
both the oracle and decoder Kraus sums, with $\gamma$ ranging over $\star$ and every complement label $\mu$. Equations~\eqref{eq:supp_face_kraus}
and \eqref{eq:supp_complete_r_block} yield
\begin{align}\label{eq:supp_face_complete_identity}
&[(\mathcal D_{ij}\otimes\operatorname{id}_{\Lop(R)})\circ
 (\mathsf S_d^{(r)}(\mathcal{I},\ldots,\mathcal A_i,\ldots,
  \mathcal A_j,\ldots,\mathcal{I})\otimes\operatorname{id}_{\Lop(R)})
 \circ(\mathcal E_{ij}\otimes\operatorname{id}_{\Lop(R)})]
 (\ket c\bra{c'}\otimes X)\notag\\
&=\sum_\gamma\sum_{a,b}
 [D_\gamma W_{a,b}(J_{ij}^{\rm in}\otimes \id_d)\otimes \id_R]
 (\ket c\bra{c'}\otimes X)
 [D_\gamma W_{a,b}(J_{ij}^{\rm in}\otimes \id_d)\otimes \id_R]^\dagger\notag\\
&=\ket c\bra{c'}\otimes\sum_{a,b}
 (V_{\pi_c,ab}\otimes \id_R)X(V_{\pi_{c'},ab}^\dagger\otimes \id_R).
\end{align}
All complement decoder terms vanish. The remaining sum has the same
$a,b$ in the two products, so it gives every control block of the two-channel
SWITCH and retains the entire target and reference output.

Now suppose the named QC--CC circuit $\mathcal C_{\boldsymbol K}$ implements the complete $r$-switch universally. Its equality with the ideal channel on the identity-filled inputs gives
\begin{equation}\label{eq:supp_pair_simulator}
\mathcal D_{ij}\circ\mathcal C_{\boldsymbol K}
 [\mathcal{I},\ldots,\mathcal A_i,\ldots,\mathcal A_j,\ldots,\mathcal{I}]
 \circ\mathcal E_{ij}
=\mathsf S_d^{(2)}(\mathcal A_i,\mathcal A_j).
\end{equation}
Apply Proposition~\ref{prop:extraction} to the named inventory
$\Omega_{\boldsymbol K}=\mathop{\dot\bigcup}_{\ell=1}^r
\{(\ell,n):1\leq n\leq K_\ell\}$, with
$\Xi=\{(i,n):1\leq n\leq K_i\}\mathbin{\dot\cup}
\{(j,n):1\leq n\leq K_j\}$ and the encoder $\mathcal E_{ij}$ and decoder $\mathcal D_{ij}$ just defined.
Every removed call is a known identity operation absorbed into the
compositions in Eq.~\eqref{eq:expanded-path}. Its record and quantum memory
remain in the direct sums of that proposition. The normalized reduced
instruments therefore define a QC--CC circuit, and the Kraus reassembly
in Eq.~\eqref{eq:complete-retention} identifies its complete output with
the left-hand side of Eq.~\eqref{eq:supp_pair_simulator}.
The resulting circuit makes exactly $K_i$ calls to $\mathcal A_i$ and $K_j$ calls to $\mathcal A_j$. For $r=2$ no unselected label is present,
and the same construction leaves the original inventory.
\end{proof}

\begin{theorem}[Exact universal complete-switch obstruction]\label{thm:supp_all_finite_r}
For every fixed finite $r,d\geq2$ and every fixed finite inventory
$\boldsymbol K\in\mathbb Z_{\geq0}^r$, no deterministic universal QC--CC
circuit reproduces the complete channel $\mathsf S_d^{(r)}$.
\end{theorem}
\begin{proof}
If $K_\ell=0$, fill all other inputs with identity channels and take a
permutation basis state $\ket\pi$ with target $\ket0$. The ideal output is
\begin{equation}\label{eq:supp_r_omitted}
\mathsf S_d^{(r)}(\mathcal{I},\ldots,\mathcal A_\ell,\ldots,\mathcal{I})
(\ket\pi\bra\pi\otimes\ket0\bra0)
=\ket\pi\bra\pi\otimes\mathcal A_\ell(\ket0\bra0).
\end{equation}
The choices $\mathcal{I}$ and $\mathcal U_{01}$ from
Lemma~\ref{lem:supp_omitted_binary} give orthogonal target outputs.
A circuit omitting $\ell$ has the same output for these choices and cannot
reproduce both. If all counts are positive, select any pair $i,j$.
Proposition~\ref{prop:supp_complete_pair_reduction} produces an exact
universal two-channel QC--CC simulator with finite counts $K_i,K_j$,
contradicting Theorem~\ref{thm:supp_all_finite_binary}.
The counts of the other inputs are unrestricted.
\end{proof}

Quantum control of causal order realizes the complete $r$-switch with one
call to each input, for $r$ calls in total~\cite{PRXQuantum.2.030335}.
For the same unknown memoryless CPTP input channels, no QC--CC circuit with a fixed finite query inventory
can reproduce that complete operation exactly and universally.

\bibliographystyle{apsrev4-2-compact}
\bibliography{refs-clean}
\end{document}